\documentclass[11pt,english]{article}

\newcommand{\CO}{{\cal O}}

\newcommand{\CD}{{\cal D}}

\usepackage{amsthm}

\makeatletter
\newcommand*{\rom}[1]{\expandafter\@slowromancap\romannumeral #1@}
\makeatother

\usepackage[latin9]{inputenc}
\usepackage{geometry}
\usepackage{verbatim}
\usepackage{prettyref}
\usepackage[numbers,sort&compress]{natbib}
\usepackage{amsmath}
\usepackage{amssymb}
\usepackage{tikz-cd}
\tikzset{commutative diagrams/row sep/huge=4cm}
\tikzset{commutative diagrams/column sep/huge=4cm}
\tikzcdset{scale cd/.style={every label/.append style={scale=#1},
    cells={nodes={scale=#1}}}}
\usepackage{lmodern}
\usepackage{tcolorbox}
\usepackage{cases}
\usepackage{bbold}
\usepackage{bm}
\usepackage{ytableau}
\usepackage{youngtab}
\usepackage{mathtools}
\usepackage{graphicx}
\usepackage[nottoc]{tocbibind}
\usepackage{mathrsfs}
\usepackage{caption}
\usepackage{subcaption}
\usepackage{cancel}
\usepackage{float}
\usepackage{tikz}
\tikzset{
    Witten diagram/.style={
        execute at begin picture={
            \draw[blue, line width=1.5pt] circle[radius=\pgfkeysvalueof{/tikz/Witten/radius}];
            \path node (X){\phantom{X}};
        },
        baseline={(X.base)}
    },
    vertex/.style={circle,fill,inner sep=0.8pt,node contents={}},
    Witten/.cd,
    radius/.initial=3cm
}
  \usetikzlibrary{shapes.misc}  
  \tikzset{cross/.style={cross out, draw=black, minimum size=2*(#1-\pgflinewidth), inner sep=0pt, outer sep=0pt},
cross/.default={1.5pt}}   
\usetikzlibrary{patterns}
\usepackage{lipsum}
\usepackage{setspace}
\usepackage{framed}
\usepackage{adjustbox}
\usepackage{mathtools}
\usepackage{braket}

\usepackage{slashed}
\usepackage{esvect}
\usepackage{dsfont}

\usepackage{color}
\definecolor{darkgreen}{rgb}{0,0.5,0}
\definecolor{darkblue}{rgb}{0,0,0.6}
\definecolor{purple}{rgb}{0.4,.2,0.7}
\usepackage[colorlinks=true,citecolor=darkgreen,linkcolor=purple,urlcolor=purple]{hyperref}

\numberwithin{equation}{section}
\numberwithin{figure}{section}
\numberwithin{table}{section}

\def\CG{{\cal G}}

\def\CD{{\cal D}}

\newcommand{\CA}{\mathcal{A}}

\DeclareFontShape{OT1}{cmr}{mx}{n}{<->cmr10}{}

\begin{document}

\fontseries{mx}\selectfont

\begin{center}
\LARGE \bf Sphere free energy of scalar field theories \\ with cubic interactions
\end{center}

\vskip1cm

\begin{center}
Simone Giombi,$^1$ Elizabeth Himwich,$^{2, 3}$ Andrei Katsevich,$^1$ Igor Klebanov,$^{1,2}$ and Zimo Sun$^{1, 3}$
\vskip5mm
{\it{\footnotesize $^1$ Joseph Henry Laboratories, Princeton University, Princeton, NJ 08544, USA} \\
\it{\footnotesize $^2$ Princeton Center for Theoretical Science, Princeton University, Princeton, NJ 08544, USA}\\
\it{\footnotesize $^3$ Princeton Gravity Initiative, Princeton University, Princeton, NJ 08544, USA}\\}
\end{center}

\vskip10mm
{\bf \center Abstract\\}
\vskip8mm
{The dimensional continuation approach to calculating the free energy of $d$-dimensional Euclidean CFT on the round sphere $S^d$ has been used to develop its $4-\epsilon$ expansion for a number of well-known non-supersymmetric theories, such as the $O(N)$ model. The resulting estimate of the sphere free energy $F$ in the 3D Ising model has turned out to be in good agreement with the numerical value obtained using fuzzy sphere regularization. In this paper, we develop the $6-\epsilon$ expansions for CFTs on $S^d$ described by scalar field theory with cubic interactions and use their resummations to estimate the values of $F$. In particular, we  
study the theories with purely imaginary coupling constants, which describe non-unitary universality classes arising when certain conformal minimal models are continued above two dimensions. The Yang-Lee model $M(2,5)$ is described by a field theory with one scalar field, while the $D$-series $M(3,8)$ model is described by two scalar fields. We also study the $OSp(1|2)$ symmetric cubic theory of one commuting and two anti-commuting scalar fields, which appears to describe the critical behavior of random spanning forests. In the course of our work, we revisit the calculation of beta functions of marginal operators containing the curvature.  We also use another method for approximating $F$, which relies on perturbation theory around the bilocal action near the long-range/short-range crossover. The numerical values it gives for $F$ tend to be in good agreement with other available methods.  }

\newpage

\tableofcontents{}

\section{Introduction and summary}

Studying $d$-dimensional Euclidean field theory on a round sphere $S^d$ has a number of motivations, which range from infrared regularization of massless field theories \cite{Adler:1972qq} to Euclidean continuation of Quantum Field Theory in de Sitter space \cite{Gibbons:1976ue}. Our main motivation in this paper is related to
 defining a measure of the number of degrees of freedom in Conformal Field Theory (CFT) 
\cite{Cardy:1988cwa,Jafferis:2011zi,Klebanov:2011gs}. A Euclidean CFT may be mapped from flat space $\mathbb{R}^d$ to a round sphere $S^d$ of radius $R$ using a Weyl transformation. Then the Euclidean path integral $Z_{S^d}$ is regulated in the infrared, and the sphere free energy $F_{S^d}=- \log  Z_{S^d}$ provides a very useful measure of the number of degrees of freedom. For even $d$, $F_{S^d}$ grows as $\log R$ due to the Weyl anomaly. For example, for $d=2$, $F_{S^2}= -\frac{c}{3} \log R+ \ldots$, where $c$ is the Weyl anomaly coefficient, which is the central charge of the Virasoro algebra. It provides an important measure of the number of degrees of freedom in a CFT, and for unitary theories it decreases along Renormalization Group (RG) flow;
this is the celebrated $c$ theorem \cite{Zamolodchikov:1986gt}. Analogously, in $d=4$,  $F_{S^4}= a \log R+ \ldots$, where $a$ is a Weyl anomaly coefficient; it satisfies the $a$ theorem \cite{Cardy:1988cwa,Jack:1990eb,Komargodski:2011vj}.

For odd-dimensional CFTs, there is no Weyl anomaly, so that the sphere free energy is independent of $R$. It is a non-local observable, a pure number characterizing the CFT. 
The sphere free energy
 satisfies an RG inequality, $F_{\rm UV}> F_{\rm IR}$, called the $F$-theorem \cite{Jafferis:2011zi,Klebanov:2011gs,Myers:2010xs}. Just like the $c$ and $a$ theorems, it is generally applicable only to unitary theories. A physical interpretation of $F$ is that it equals the long-range part of the Quantum Entanglement Entropy across a $d-2$ dimensional sphere \cite{Casini:2011kv}. This connection with the Entanglement Entropy has enabled a proof of the $F$-theorem in the $d=3$ case \cite{Casini:2012ei}.

The calculation of the sphere free energy $F$ in some 3D CFTs with extended supersymmetry may be reduced to finite dimensional integrals using the methods of localization \cite{Pufu:2016zxm}. For the well-known non-supersymmetric CFTs, such as the $O(N)$ vector model, the Gross-Neveu-Yukawa theory, and conformal QED, one instead has to resort to approximate methods for calculating $F$.
They include dimensional continuation of the smooth quantity $\widetilde{F} = - \sin\left(\frac{\pi d}{2}\right)F$
\cite{Giombi:2014xxa,Fei:2015oha,Giombi:2015haa,Fei:2016sgs,Tarnopolsky:2016vvd,DeCesare:2022obt} and the $1/N$ expansion \cite{Klebanov:2011gs,Klebanov:2011td,Tarnopolsky:2016vvd,Fraser-Taliente:2025udk}. More recently, a novel numerical method for studying 3D CFTs was introduced 
\cite{Zhu:2022gjc}, and it is attracting considerable interest (see also \cite{Dedushenko:2024nwi,Zhou:2024dbt} for applications of this approach to boundary CFTs).
This method relies on mapping the spectrum of operator dimensions to the energy spectrum on the spatial 2-sphere supplied with a ``fuzzy sphere" regulator. 
For the case of the 3D Ising model, the fuzzy sphere approach has produced excellent agreement with other calculations of the spectrum of scaling dimensions. This method can also be applied to the calculation of $F$, since it may be defined as the Entanglement Entropy across the equator of the spatial sphere. The numerical evaluation of this quantity for the 3D Ising model 
\cite{Hu:2024pen}
produced excellent agreement with the earlier result found via dimensional continuation \cite{Giombi:2014xxa,Fei:2015oha}. This success motivates us to revisit the dimensional continuation approach to calculating $F$ and extend it to additional conformal models that are non-unitary.

We will also use another method for approximating $F$, which we call the Long-Range Approach (LRA). In this approach, one begins with the free long-range bilocal action (\ref{S0LR}), adjusts its parameter $s$ to match the field dimension in the usual short-range CFT, and subsequently includes the interaction corrections perturbatively. We find that the LRA tends to agree well with the dimensional continuation approach, and in some cases may work even better.

One of the topics of this paper is the sphere free energy for theories of $N+1$ scalar fields $\phi^i, i=1, \ldots N$ and $\sigma$ with cubic interactions 
$\frac{1}{2} g_1 \sigma \phi^i \phi^i+ \frac{1}{6} g_2 \sigma^3$. When $N$ is sufficiently large, these theories possess real perturbative fixed points that describe continuation of the $O(N)$ model above $4$ dimensions \cite{Fei:2014yja}. The $6-\epsilon$ expansion of this theory has been studied at higher orders  \cite{Fei:2014xta,Gracey:2015tta,Kompaniets:2021hwg}, and the $1/N$ expansion of the sphere free energy was presented in \cite{Tarnopolsky:2016vvd}. 
In this paper, we instead focus on the fixed points with purely imaginary coupling constants, which are present for $N\leq 1$  \cite{Fei:2014xta}.\footnote{For scalar field theories with $O(N)^3$ symmetry \cite{Klebanov:2016xxf,Giombi:2017dtl}, there are large $N$ melonic fixed points with imaginary tetrahedral coupling constant \cite{Benedetti:2019eyl}. The sphere free energy in melonic theories 
was studied in \cite{Benedetti:2021wzt,Fraser-Taliente:2024hzv}.
}
 The earliest and best-known example is $N=0$; it is the theory of one scalar field with the $i\sigma^3$ interaction, which describes the Yang-Lee universality class
\cite{PhysRevLett.40.1610}. When continued from $6-\epsilon$ to $2$ dimensions, it becomes equivalent \cite{PhysRevLett.54.1354} to the $M(2,5)$ non-unitary conformal minimal model 
\cite{Belavin:1984vu} with central charge $c(2,5)=-\frac{22}{5}$. A similar theory with $N=1$ \cite{Fei:2014xta}, which has been receiving some attention recently \cite{Klebanov:2022syt,Nakayama:2024msv,Katsevich:2024jgq,Delouche:2024tjf}, describes  
the universality class of the $D$-series $M(3,8)$ non-unitary minimal model; its central charge is $c(3,8)=-\frac{21}{4}$. 
We will also consider the formal continuation to $N=-2$, which is described by one commuting and two anti-commuting scalar fields (a pair of symplectic fermions). Such a theory has an $OSp(1|2)$ 
invariant fixed point in $6-\epsilon$ dimensions \cite{Fei:2015kta,Klebanov:2021sos}. It is equivalent to the formal $q\rightarrow 0$ limit of the critical $q$-state Potts model, which describes random spanning forests. This non-unitary statistical model, which is critical in $2<d<6$ dimensions, has been studied using various mathematical and numerical methods \cite{Caracciolo:2004hz,Deng:2006ur,Bauerschmidt:2019mhu}.

Using dimensional continuation methods similar to those in \cite{Giombi:2014xxa,Fei:2015oha,Tarnopolsky:2016vvd}, we will study the sphere free energies of the Yang-Lee, $M(3,8)$, and
$OSp(1|2)$ universality classes. The necessary Feynman diagrams can be evaluated using Mellin-Barnes integrals with the \textit{Mathematica} packages {\fontfamily{qhv}\selectfont\small MB.m} \cite{Czakon:2005rk,Smirnov:2009up,Smirnov:2012gma,Belitsky:2022gba} as described in Appendix B of \cite{Fei:2015oha}, to which we refer readers for details. 
The $6-\epsilon$ expansion of $\widetilde F$ is affected by the existence of nearly marginal curvature terms $\mathcal{R}^3$, $ \mathcal{R}^2 \sigma$, $\mathcal{R} \sigma^2$, and 
$ \mathcal{R} \phi^i \phi^i$. We calculate the beta functions for these terms, and find results differing from some of the earlier literature. 
We find that, with the exception of $\mathcal{R}^3$, the curvature terms do not affect the expansion of $\widetilde F$ up to order $\epsilon^3$.

This paper is organized as follows. In Section \ref{sec:ON}, we consider renormalization of the $O(N)$ cubic model. We begin with a flat-space warmup/review in Section \ref{sec:warmup} before turning to the renormalization of curvature couplings in Section \ref{sec:curvren}. In Section \ref{sec:free}, we discuss the sphere free energy and its renormalization in dimensional regularization.  In Section \ref{sec:NL} we consider the calculation of the sphere free energy using the long-range approach, with both quartic (Section \ref{sec:quarticLR}) and cubic (Section \ref{sec:cubicLR}) interactions. In Section \ref{sec:numerics} we discuss numerical results for the free energy, including its Pad\'{e} approximant for cubic $N=0,1,-2$ models in dimensional regularization and a comparison with results from long-range models.  Appendix \ref{app:flat} reviews renormalization of the $O(N)$ cubic model in flat space, Appendix \ref{app:conventions} summarizes our conventions for integrals on the sphere, Appendix \ref{app:details} includes details of the free energy calculation in dimensional regularization and an example integral evaluation, Appendix \ref{app:comparison} compares our results for cubic $N=0$ with previous literature, and Appendix \ref{app:NL} includes details of the free energy calculations using the long-range approach.

\section{Cubic scalar $O(N)$  theory} \label{sec:ON}

In this section, we consider the renormalization of the $O(N)$ cubic model on a sphere in $d=6-\epsilon$ dimensions. We first introduce the model and review its renormalization in flat space, and then discuss the renormalization of curvature counterterms.

\subsection{Flat-space warmup} \label{sec:warmup}

The flat-space cubic $O(N)$ theory in $d=6-\epsilon$ dimensions was introduced in \cite{Fei:2014yja}, and its renormalization was further studied in \cite{Fei:2014xta,Fei:2015kta,Gracey:2015tta,Giombi:2019upv,Arias-Tamargo:2020fow,Kompaniets:2021hwg,Antipin:2021jiw,Jack:2021ziq}.\footnote{Early studies of the $N=0$ case in $d=6-\epsilon$ were performed in \cite{MACFARLANE197491,PhysRevLett.40.1610,deAlcantaraBonfim:1980pe,OFdeAlcantaraBonfirm_1981,Barbosa:1986kv}. Studies of this model have also been performed using the functional renormalization group \cite{Mati:2014xma,Mati:2016wjn,Eichhorn:2016hdi,Kamikado:2016dvw,Zambelli:2016cbw} and conformal bootstrap \cite{Chester:2014gqa,Li:2016wdp}.  } The bare action for the $O(N)$-symmetric model of $N+1$ scalar fields in flat space is 
\begin{equation}\label{cubicO(N)flat}
  \begin{aligned}
    S&=\int d^dx\Bigg(\frac{1}{2}(\partial_\mu\phi_0^i)^2+\frac{1}{2}(\partial_\mu\sigma_0)^2+\frac{1}{2}g_{1,0}\sigma_0\phi_0^i\phi_0^i+\frac{1}{3!}g_{2,0}\sigma_0^3 \Bigg),
    \end{aligned}
\end{equation}
where $\sigma_0$ and $\phi^i_0$, $i = 1, ... , N$ are bare scalar fields and $g_{1,0}$, $g_{2,0}$ are bare couplings with classical scaling dimension $\epsilon/2$ in $d=6-\epsilon$. To renormalize the theory, one introduces renormalized fields and coupling constants \begin{equation} \label{eq:Zdef_flat}
\begin{aligned}
    &\phi^i_0=Z_\phi^{\frac{1}{2}}\phi^i,\quad\sigma_0=Z_\sigma^{\frac{1}{2}}\sigma,\quad g_{1,0}=\mu^{\frac{\epsilon}{2}}Z_\phi^{-1}Z_\sigma^{-\frac{1}{2}}Z_{g_1}g_1,\quad g_{2,0}=\mu^{\frac{\epsilon}{2}}Z_\sigma^{-\frac{3}{2}}Z_{g_2}g_2,\\
\end{aligned}
\end{equation}
where $\mu$ is an auxiliary scale and 
\begin{equation} \label{eq:ctdef_flat}
\begin{aligned}
    &Z_\phi=1+\delta_\phi,\quad Z_\sigma=1+\delta_\sigma,\quad Z_{g_1}=1+\frac{\delta_{g_1}}{g_1},\quad Z_{g_2}=1+\frac{\delta_{g_2}}{g_2}.
\end{aligned}
\end{equation}
In this section, we use dimensional regularization \cite{tHooft:1972tcz} in $d=6-\epsilon$ and the minimal subtraction renormalization scheme \cite{Hooft1973DimensionalRA}. A summary of previous results for flat-space renormalization are collected for reference in Appendix \ref{app:flat}.

As a warmup example of our method, it is instructive to review the scalar two-point function renormalization. While the momentum-space representation of Feynman diagrams is very helpful in flat space, there is no simple analog on the sphere, where we instead work in position space. When moving from momentum space to position space, it is important to keep track of reducible diagrams. For instance, consider the momentum-space two-point function of scalar fields $\braket{\varphi_0(p)\varphi_0(-p)}$, where $\varphi_0$ is either $\sigma_0$ or $\phi^i_0$, and $\varphi_0 = Z_{\varphi}^{1/2} \varphi$. Defining $\Sigma(p^2)$ as the sum of all one-particle-irreducible insertions into the propagator, the full two-point function in momentum-space is given by the geometric series
\begin{equation} \label{eq:twoptexpansion}
\braket{\varphi_0(p)\varphi_0(-p)} = \frac{1}{p^2} + \frac{1}{p^2}\Sigma(p^2)\frac{1}{p^2} + \frac{1}{p^2}\Sigma(p^2)\frac{1}{p^2}\Sigma(p^2)\frac{1}{p^2} + \cdots = \frac{1}{p^2 - \Sigma(p^2)}. 
\end{equation}
In turn, $\Sigma(p^2)$ can be expanded as a sum over $k$-loop diagrams denoted $\Sigma_k(p^2)$, which each have $2k$ powers of the cubic coupling $g_0$:
\begin{equation}
\Sigma(p^2) = g_0^2~ \Sigma_1(p^2) + g_0^4~ \Sigma_2(p^2) + \cdots. 
\end{equation}
One often discusses momentum-space renormalization conditions on the one-particle-irreducible diagrams $\Sigma(p^2)$. However, it is important to remember that when renormalizing the two-point function by requiring that $Z_\varphi^{-1}\langle \varphi_0(p)\varphi_0(-p)\rangle$ is finite order-by-order in $g^2$, reducible diagrams will appear in the expansion \eqref{eq:twoptexpansion}. For instance, at $\mathcal{O}(g^4)$, the irreducible diagrams $\frac{1}{p^2}\Sigma_2(p^2)\frac{1}{p^2}$ as well as the reducible diagrams $\frac{1}{p^2}\Sigma_1(p^2)\frac{1}{p^2}\Sigma_1(p^2)\frac{1}{p^2}$ will contribute. The same diagrams appear in the position-space calculation, which is  equivalent.

\begin{figure}[t]
\centering
\begin{tikzpicture} 
\draw[color=black] (0,0) circle [radius=0.6]; 
\draw (-1.3, 0) to  (-0.6,0)node[vertex]{} ;
\draw (1.3, 0) to  (0.6,0)node[vertex]{} ;
\node at (0,-1.1) {$\CG_2$};
\end{tikzpicture} 
\quad
\begin{tikzpicture} 
\draw[color=black] (0,0) circle [radius=0.6]; 
\draw (-1.3, 0) to  (-0.6,0)node[vertex]{} ;
\draw (1.3, 0) to  (0.6,0)node[vertex]{} ;
\draw (0, 0.6)node[vertex]{} to  (0,-0.6)node[vertex]{} ;
\node at (0,-1) {$\CG_4^{(1)}$};
\end{tikzpicture} 
\quad
\begin{tikzpicture} 
\draw[color=black] (0,0) circle [radius=0.6]; 
\draw (-1.3, 0) to  (-0.6,0)node[vertex]{} ;
\draw (1.3, 0) to  (0.6,0)node[vertex]{} ;
\draw (-0.4, 0.45)node[vertex]{} to  (0.4,0.45)node[vertex]{} ;
\node at (0,-1) {$\CG_4^{(2)}$};
\end{tikzpicture} \\
\centering
\begin{tikzpicture} 
\draw[color=black] (0,0) circle [radius=0.6]; 
\draw[color=black] (1.9,0) circle [radius=0.6]; 
\draw (-1.3, 0) to  (-0.6,0)node[vertex]{} ;
\draw (1.3, 0)node[vertex]{} to  (0.6,0)node[vertex]{} ;
\draw (2.5, 0)node[vertex]{} to  (3.2,0);
\node at (1,-1) {$\CG_4^{(3)}$};
\end{tikzpicture} 
\quad
\begin{tikzpicture} 
\draw[color=black] (0,0) circle [radius=0.6]; 
\draw (-0.6, 0) node[vertex]{} to (0.6, 0)node[vertex]{}; 
\draw (0, -0.6) node[vertex]{} to (0, -0.8)node[vertex]{}; 
\draw (-1.3, -0.8) to (1.3, -0.8); 
\node at (0,-1.2) {$\CG_4^{(4)}$};
\end{tikzpicture} 
\caption{One and two-loop corrections to the propagator to order $\CO(g_1^{n_1}g_2^{n_2})$ with $n_1+n_2=4$.}
\label{phiphi}
\end{figure}
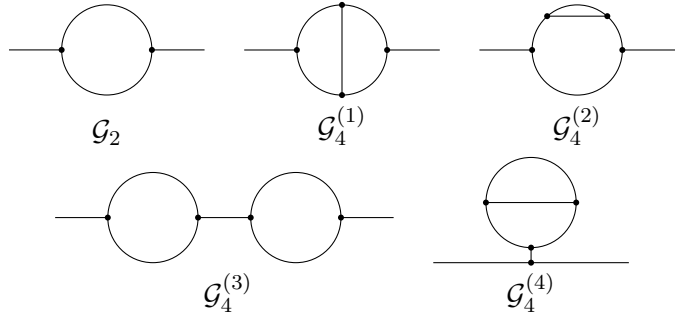
In the cubic scalar $O(N)$ theory, the flat position-space two-point functions to order $\CO(g_1^{n_1}g_2^{n_2})$ with $n_1+n_2=4$ are (see Figure \ref{phiphi}) 
\begin{equation}
\begin{aligned}
    \braket{\sigma_0(x)\sigma_0(y)}&= \mathbb{G}_d(x,y) \Bigg(1 +\frac{Ng^2_{1,0}+g^2_{2,0}}{2}\mathcal{G}_2 
    +\frac{Ng^4_{1,0} + 2Ng^3_{1,0}g_{2,0} + g_{2,0}^4}{2}\mathcal{G}^{(1)}_4 \\ &\qquad \qquad \qquad + \frac{2Ng_{1,0}^4 + Ng^2_{1,0}g^2_{2,0}+ g_{2,0}^4}{2}\mathcal{G}^{(2)}_4 + \frac{N^2g_{1,0}^4 + 2Ng^2_{1,0}g^2_{2,0} + g_{2,0}^4}{4}\mathcal{G}^{(3)}_4 \\
    &\qquad \qquad \qquad +\frac{2Ng^3_{1,0}g_{2,0}+ Ng^2_{1,0}g^2_{2,0}+ g_{2,0}^4}{4}\mathcal{G}^{(4)}_4 + \cdots \Bigg),
\end{aligned}
\end{equation}
\begin{equation}
\begin{aligned}
    \braket{\phi^i_0(x)\phi^j_0(y)}&=\delta^{ij}\mathbb{G}_d(x,y)\Bigg(1+g^2_{1,0}\mathcal{G}_2
    +\left(g^4_{1,0}+g^3_{1,0}g_{2,0}\right)\mathcal{G}^{(1)}_4+\frac{(N+2)g^4_{1,0}+g_{1,0}^2g_{2,0}^2}{2}\mathcal{G}^{(2)}_4 \\ &\qquad \qquad \qquad \qquad +g^4_{1,0}\mathcal{G}^{(3)}_4 +\frac{2Ng^4_{1,0} + Ng^3_{1,0}g_{2,0} + g_{1,0}g_{2,0}^3}{4}\mathcal{G}^{(4)}_4 + \cdots \Bigg),
\end{aligned}
\end{equation}
where we have pulled out an overall factor of the massless propagator $\mathbb{G}_d(x,y)$ defined in \eqref{eq:flatprop}, and the dots denote $\CO(g_1^{n_1}g_2^{n_2})$ with $n_1+n_2 = 6$. Using Mellin-Barnes integrals \cite{Czakon:2005rk,Smirnov:2009up,Smirnov:2012gma,Belitsky:2022gba} to evaluate the diagrams, we find 
\begin{equation} \label{eq:flatphiphi}
  \begin{aligned}
    \mathcal{G}_2&=-\frac{1}{3(4\pi)^3}\left(\frac{1}{\epsilon}+\frac{5/3+\log(\pi e^{\gamma_E}|x-y|^2)}{2}+\mathcal{O}(\epsilon)\right), \\
    \mathcal{G}^{(k)}_4&=-\frac{1}{3(4\pi)^6}\begin{cases}\frac{1}{\epsilon^2}+\frac{2+\log(\pi e^{\gamma_E}|x-y|^2)}{\epsilon}+\mathcal{O}(1), &k=1, \\
    -\frac{1}{6}\left(\frac{1}{\epsilon^2}+\frac{31+12\log(\pi e^{\gamma_E}|x-y|^2)}{12\epsilon}\right) +\mathcal{O}(1), &k=2, \\
    -\frac{1}{3}\left(\frac{1}{\epsilon^2}+\frac{5+3\log(\pi e^{\gamma_E}|x-y|^2)}{3\epsilon}\right) +\mathcal{O}(1), &k=3, \\
    %\mathcal{O}(1), &k=4, \\
    \end{cases}
    \end{aligned}
\end{equation}
where $\gamma_E$ is the Euler-Mascheroni constant. The diagram $\mathcal{G}_4^{(4)}$ is $\mathcal{O}(\epsilon)$.  When the two-point function is expressed in terms of the renormalized couplings via \eqref{eq:Zdef_flat}, additional factors of $\log \mu$ will make the argument of the subleading logarithm dimensionless. Combining all the diagrams and requiring that $Z_\phi^{-1}\langle \phi_0^i(x)\phi_0^j(y)\rangle$ and $Z_\sigma^{-1}\langle \sigma_0(x)\sigma_0(y)\rangle$ are finite, we recover the wavefunction renormalization \eqref{eq:Zphi} and \eqref{eq:Zsigma}. Note that the reducible diagram $\mathcal{G}_4^{(3)}$ in Fig.~\ref{phiphi} must be included to obtain the correct result.

\subsection{Renormalization on the sphere} \label{sec:curvren}

The generalization of the cubic scalar $O(N)$  model to the sphere $S^{6-\epsilon}$ was considered in \cite{Giombi:2014xxa} and \cite{Tarnopolsky:2016vvd}, which respectively computed the leading and subleading terms in the $\epsilon$-expansion of the sphere free energy. To renormalize the theory in curved space, one must consider all independent consistent curvature couplings that are nearly marginal in $d=6-\epsilon$, following \cite{Brown:1980qq,Hathrell:1981zb,Toms:1982af,JACK1986139}. On the sphere, we thus have the following bare action for the $O(N)$-symmetric cubic model of $N+1$ scalar fields:
\begin{equation}\label{cubicO(N)}
  \begin{aligned}
    S&=\int d^dx\sqrt{g}\Bigg(\frac{1}{2}(\partial_\mu\phi_0^i)^2+\frac{1}{2}(\partial_\mu\sigma_0)^2+\frac{\xi}{2}\mathcal{R}\left(\phi_0^i\phi_0^i+\sigma_0^2\right)+\frac{1}{2}g_{1,0}\sigma_0\phi_0^i\phi_0^i+\frac{1}{3!}g_{2,0}\sigma_0^3 \\&\qquad \qquad \qquad \qquad+\frac{\eta_{1,0}}{2}\mathcal{R}\phi_0^i\phi_0^i+\frac{\eta_{2,0}}{2}\mathcal{R}\sigma_0^2+\kappa_0\mathcal{R}^2\sigma_0+b_0\mathcal{R}^3\Bigg),
    \end{aligned}
\end{equation}
where $\mathcal{R}$ is the Ricci scalar curvature, $\sigma_0$ and $\phi^i_0$, $i = 1, ... , N$ are bare conformally coupled scalar fields with $\xi=\frac{d-2}{4(d-1)}$, and $g_{1,0}$, $g_{2,0}$, $\eta_{1,0}$, $\eta_{2,0}$, $\kappa_0$, and $b_0$ are bare couplings. Note that we have separated arbitrary couplings $\eta_{1,0}$ and $\eta_{2,0}$ from the conformal coupling term $\xi$, and that a linear term in $\phi^i_0$ is forbidden by $O(N)$ symmetry. In $d=6-\epsilon$ dimensions, $\kappa$ has classical scaling dimension $-\epsilon/2$, $\eta$ has dimension zero, and $b$ has dimension $-\epsilon$. 
Our conventions for propagators and integrals on the sphere are collected in Appendix \ref{app:conventions}.

Following the definitions \eqref{eq:Zdef_flat} and \eqref{eq:ctdef_flat} in flat space, on the sphere we introduce renormalized fields and coupling constants 
\begin{equation} \label{eq:baresphere}
\begin{aligned}
    &\phi^i_0=Z_\phi^{\frac{1}{2}}\phi^i,\quad\sigma_0=Z_\sigma^{\frac{1}{2}}\sigma,\quad g_{1,0}=\mu^{\frac{\epsilon}{2}}Z_\phi^{-1}Z_\sigma^{-\frac{1}{2}}Z_{g_1}g_1,\quad g_{2,0}=\mu^{\frac{\epsilon}{2}}Z_\sigma^{-\frac{3}{2}}Z_{g_2}g_2,\\
    &\eta_{1,0}=Z_\phi^{-1}Z_{\eta_1}\eta_1,\quad \eta_{2,0}=Z_\sigma^{-1}Z_{\eta_2}\eta_2,\quad\kappa_0=\mu^{-\frac{\epsilon}{2}}Z_\sigma^{-\frac{1}{2}}Z_\kappa\kappa,\quad b_0=\mu^{-\epsilon}Z_bb,
\end{aligned}
\end{equation}
where $\mu$ is an auxiliary scale and 
\begin{equation} \label{eq:ctdef}
\begin{aligned}
    &Z_\phi=1+\delta_\phi,\quad Z_\sigma=1+\delta_\sigma,\quad Z_{g_1}=1+\frac{\delta_{g_1}}{g_1},\quad Z_{g_2}=1+\frac{\delta_{g_2}}{g_2},\\
    &Z_{\eta_1}=1+\frac{\delta_{\eta_1}}{\eta_1},\quad Z_{\eta_2}=1+\frac{\delta_{\eta_2}}{\eta_2},\quad Z_\kappa=1+\frac{\delta_\kappa}{\kappa},\quad Z_b=1+\frac{\delta_b}{b}.
\end{aligned}
\end{equation}
The renormalization of $g_1$ and $g_2$ on the sphere is the same as in flat space, and was computed in \cite{Fei:2014yja,Fei:2014xta,Gracey:2015tta} (see Appendix \ref{app:flat} for a summary).  For the renormalization of the sphere free energy to order $\epsilon^2$ in Section \ref{sec:free}, we need the relation between the bare and renormalized $g_1$ and $g_2$ couplings to quartic order, which are shown in \eqref{eq:g10} and \eqref{eq:g20}.

The next-to-leading term in the sphere free energy for large $N$ was computed in \cite{Tarnopolsky:2016vvd}, to which order the contribution of $\eta_1$, $\eta_2$, and $\kappa$ was assumed to vanish. The renormalization of $b$ was then determined by finiteness of the renormalized sphere free energy to sixth order in $g_1$ and $g_2$, which was also assumed to be independent of $\eta_1$, $\eta_2$, and $\kappa$.  Here, we determine the renormalization of $\eta_1$, $\eta_2$, and $\kappa$ directly from finiteness of the renormalized one- and two-point functions of $\phi$ and $\sigma$.  A priori, renormalization of the curvature couplings can depend on any of the marginal couplings involving $\phi$ or $\sigma$ in the action. The curvature counterterms are expanded
\begin{equation} \label{eq:ctexpansion}
  \delta_{\eta_1,\eta_2}=\sum_{n=1}^\infty\frac{\delta^{(n)}_{\eta_1,\eta_2}}{\epsilon^n},\quad\delta_\kappa=\sum_{n=1}^\infty\frac{\delta^{(n)}_\kappa}{\epsilon^n},\quad\delta_b=\sum_{n=1}^\infty\frac{\delta^{(n)}_b}{\epsilon^n}.
\end{equation}
Using our results, we will see in Section \ref{sec:free} that the renormalized curvature couplings do not contribute to the renormalized free energy at the fixed point at order $\epsilon^2$, justifying the assumptions of \cite{Tarnopolsky:2016vvd}.

\subsubsection{One-point function}

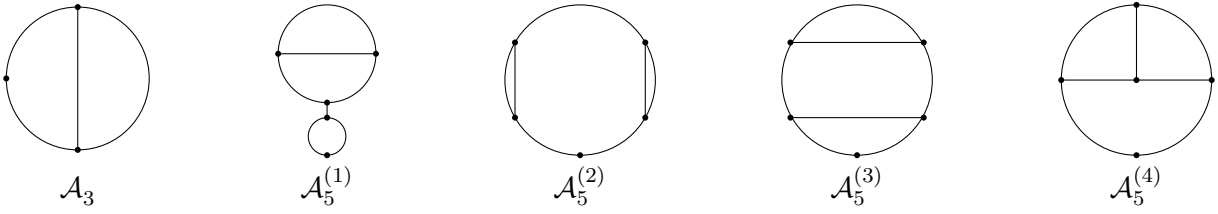
\begin{figure}[t]
\centering
\begin{tikzpicture} 
\draw[color=black] (0,0) circle [radius=0.95]; 
\draw (-0.95, 0)node[vertex]{} ;
\draw (0, 0.95)node[vertex]{} to  (0,-0.95)node[vertex]{} ;
\node at (0,-1.5) {$\mathcal{A}^{}_3$};
\end{tikzpicture} 
\qquad
\qquad
\begin{tikzpicture} 
\draw[color=black] (0,0.35) circle [radius=0.65]; 
\draw (-.65, 0.35)node[vertex]{} to  (.65,0.35)node[vertex]{} ;
\draw (0, -0.3)node[vertex]{} to  (0,-.5)node[vertex]{} ;
\draw[color=black] (0,-.75) circle [radius=0.25]; 
\draw (0, -1)node[vertex]{};
\node at (0,-1.4) {$\mathcal{A}_5^{(1)}$};
\end{tikzpicture} 
\qquad
\qquad
\begin{tikzpicture} 
\draw[color=black] (0,0) circle [radius=1]; 
\draw (-.86603, .5)node[vertex]{} to  (-.86603, -.5)node[vertex]{} ;
\draw (.86603, -.5)node[vertex]{} to  (.86603, .5)node[vertex]{} ;
\draw (0, -1)node[vertex]{} ;
\node at (0,-1.4) {$\mathcal{A}_5^{(2)}$};
\end{tikzpicture} 
 \qquad
 \qquad
\begin{tikzpicture} 
\draw[color=black] (0,0) circle [radius=1]; 
\draw (-.88603, .5)node[vertex]{} to  (.88603, .5)node[vertex]{} ;
\draw (-.88603, -.5)node[vertex]{} to  (.88603, -.5)node[vertex]{} ;
\draw (0, -1)node[vertex]{} ;
\node at (0,-1.4) {$\mathcal{A}_5^{(3)}$};
\end{tikzpicture}
\qquad 
\qquad
\begin{tikzpicture} 
\draw[color=black] (0,0) circle [radius=1]; 
\draw (-1, 0)node[vertex]{} to  (0, 0)node[vertex]{} to (1, 0)node[vertex]{} ;
\draw (0, 0) to  (0,1)node[vertex]{} ;
\draw (0, -1)node[vertex]{} ;
\node at (0,-1.4) {$\mathcal{A}_5^{(4)}$};
\end{tikzpicture} 
\caption{Diagrams up to order $\CO(g_1^{n_1}g_2^{n_2})$ with $n_1+n_2=5$ that contribute to the one-point function of $\sigma$. The external leg is amputated.} 
\label{phi5g}
\end{figure}

Here, we compute the one-point function $\braket{\sigma_0}$ on the sphere and renormalize $\braket{\sigma} = Z^{-\frac{1}{2}}_\sigma \braket{\sigma_0}$. Note that the one-point function is position-independent due to $SO(d+1)$ symmetry. First, we consider the contributions purely from $g_1$ and $g_2$. At order $\CO(g_1^{n_1}g_2^{n_2})$ with $n_1+n_2=1$, the tadpole diagram vanishes in dimensional regularization. At order $\CO(g_1^{n_1}g_2^{n_2})$ with $n_1+n_2=3$, there is only one diagram, i.e. $\mathcal{A}_3$ in Figure \ref{phi5g}. At order $\CO(g_1^{n_1}g_2^{n_2})$ with $n_1+n_2=5$, the $\CA_5^{(a)}$ diagrams in Figure \ref{phi5g} contribute. Altogether, these diagrams yield 
\begin{align} \label{eq:sigma0}
    \braket{\sigma_0} = -\left(a_3 \mathcal{A}_3+ a_{5,1}\mathcal{A}^{(1)}_5+a_{5,2}\mathcal{A}_5^{(2)}+a_{5,3}\mathcal{A}_5^{(3)}+ a_{5,4}\mathcal{A}_5^{(4)}\right)\frac{C_d I_2(\frac{d-2}{2})}{\text{Vol}(S^d)} + \cdots,
\end{align}
where the constants  ${\text{Vol}(S^d)}$ and $C_d$ are defined in \eqref{eq:vold} and \eqref{eq:flatprop}, the integral $I_2$ is defined in \eqref{eq:I2}, the factor $\frac{C_d I_2(\frac{d-2}{2})}{\text{Vol}(S^d)} = \int d^d y~\Omega^d(y)G_d(x, y)$ comes from the external leg, the dots denote contributions from curvature counterterms (considered below) as well as $\CO(g_1^{n_1}g_2^{n_2})$ with $n_1+n_2=7$, and the combinatorial factors for each diagram are
\begin{equation}
  \begin{aligned}
    a_3 &= \frac{1}{4}\left(2Ng_{1,0}^3 + Ng_{1,0}^2g_{2,0} + g_{2,0}^3\right), \\
    a_{5,k} &= \begin{cases}\frac{1}{8}\left(2 N^2g_{1,0}^5 +  N^2g_{1,0}^4g_{2,0} + 2 Ng_{1,0}^3g_{2,0}^2 + 2 Ng_{1,0}^2g_{2,0}^3 + g_{2,0}^5\right), &k=1, \\
     \frac{1}{8}\left(4 Ng_{1,0}^5 + N^2g^4_{1,0}g_{2,0} + 2 Ng_{1,0}^2g_{2,0}^3 + g_{2,0}^5\right), &k=2, \\
    \frac{1}{4}\left(N(N+2)g_{1,0}^5 + 2 Ng^4_{1,0}g_{2,0} + Ng_{1,0}^3g_{2,0}^2 + Ng_{1,0}^2g_{2,0}^3 + g_{2,0}^5\right),&k=3, \\
    \frac{1}{4}\left(2 Ng_{1,0}^5 + 3Ng^4_{1,0}g_{2,0} + 2 Ng_{1,0}^3g_{2,0}^2+ g_{2,0}^5\right), &k=4. \\
    \end{cases}
  \end{aligned}  
\end{equation}
The evaluation of $\CA_3$ is presented as an analytic example in Appendix \ref{app:analyticA}.  We find 
\begin{equation}
\begin{aligned}
    \mathcal{A}_3 &= (2R)^{8-2d}C_d^4e^{\gamma_E(d-6)+\frac{\pi^2}{24}(d-6)^2}\pi^{d}
    \left(-\frac{1}{8}-\frac{11\epsilon}{32}+\mathcal{O}(\epsilon^2)\right),\\
    \mathcal{A}_5^{(k)}&=(2R)^{14-3d}C_d^7e^{2\gamma_E(d-6)+\frac{\pi^2}{24}(d-6)^2}\pi^{2d}\begin{cases}
        \frac{1}{24\epsilon}+\frac{25}{144}+\mathcal{O}(\epsilon), &k=1,\\
        \frac{1}{12\epsilon}+\frac{193}{432}+\mathcal{O}(\epsilon), &k=2,\\
        -\frac{1}{12\epsilon}-\frac{13}{48}+\mathcal{O}(\epsilon), &k=3,\\
        -\frac{1}{4\epsilon}-\frac{73}{48}+\mathcal{O}(\epsilon), &k=4,
    \end{cases}
    \end{aligned}
\end{equation}
where $R$ is the radius of the sphere.

To renormalize the one-point function we impose that $\braket{\sigma}=Z_{\sigma}^{-1/2}\braket{\sigma_0}$ is finite. Using \eqref{eq:sigma0} and the flat-space renormalization of $\sigma$ \eqref{eq:Zsigma}, we find that the pure $g_1$ and $g_2$ contributions are finite up to $\CO(g_1^{n_1}g_2^{n_2})$ with $n_1+n_2=5$. This implies that such terms do not contribute to the renormalization of $\kappa$. 
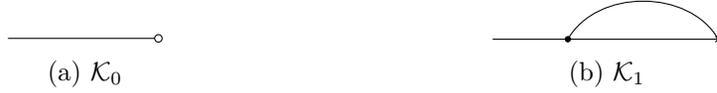
\begin{figure}[t] 
\centering
\begin{subfigure}[b]{0.4\textwidth}
\centering
\begin{tikzpicture} [ circ/.style={shape=circle, inner sep=1pt, draw, fill=white, node contents=}]
\draw (-2, 0) to  (0,0)node[circ]{} ;
\end{tikzpicture}
\caption{$\mathcal{K}_0$}
\end{subfigure}
\hspace{0.2em}
\begin{subfigure}[b]{0.4\textwidth}
\centering
\begin{tikzpicture}
\draw (-1, 0) to  (0,0)node[vertex]{} to [out=60,in=120]   (2,0)node[cross]{} to   (0,0)node[vertex]{} ;
\end{tikzpicture}
\caption{$\mathcal{K}_1$}
\end{subfigure}
\caption{Leading contributions of curvature terms to the one-point function. The empty dot represents an insertion of $\kappa$ and the cross represents an $\eta$ vertex.}
\label{fig:onept}
\end{figure}

Next, we consider the following leading contributions from diagrams with curvature couplings shown in Figure \ref{fig:onept}:
\begin{equation}
\braket{\sigma_0} \supset - \left(\kappa_0\mathcal{R}^2 \mathcal{K}_0 - \frac{Ng_{1,0}\eta_{1,0}+g_{2,0}\eta_{2,0}}{2}\mathcal{R}\mathcal{K}_1 \right)\frac{C_d I_2(\frac{d-2}{2})}{\text{Vol}(S^d)},
\end{equation}
where
\begin{equation}
\mathcal{K}_0 = 1, \ \ \ \mathcal{K}_1 = \frac{C_d^2 I_2(d-2)}{\text{Vol}(S^d)} = - \frac{2}{(4\pi)^3 R^2 \epsilon} + \mathcal{O}(1). 
\end{equation}
Since $Z_{\sigma}$ begins at $\CO(g_1^{n_1}g_2^{n_2})$ with $n_1+n_2=2$ (see \eqref{eq:Zsigma}), these leading curvature contributions imply that for $\braket{\sigma}=Z_{\sigma}^{-1/2}\braket{\sigma_0}$ to be finite, we must have 
\begin{equation} \label{eq:oneptct}
    \delta^{(1)}_\kappa=-\frac{N\eta_1 g_1+\eta_2g_2}{30(4\pi)^3}+ \cdots, 
\end{equation}
where we used the value of $\mathcal{R}$ in \eqref{eq:Ricci} and the dots denote $\mathcal{O}(g_1^{n_1}g_2^{n_2}) $ with $n_1 + n_2 = 7$ as well as curvature contributions beginning at $\mathcal{O}(\eta^2g, \eta g^3, \kappa g^4)$. Note that we do not need to include mixing of $\sigma$ with $\mathcal{R}$ for renormalization at this order, though it can contribute at higher orders.

\subsubsection{Two-point function}

Next, we consider the renormalization of two-point functions of scalar fields.  Again, we first consider the contribution of purely $g_{1,0}$ and $g_{2,0}$, and then consider contributions from the curvature vertices.

To generalize the calculations of Section \ref{sec:warmup} to the sphere, we make the choice of special points $x^2 = 1$ and $y=0$. Computing these diagrams on the sphere, we find agreement with the flat space result \eqref{eq:flatphiphi} except with the flat space distance $|x-y|^2$ replaced by the corresponding distance $D(1,0)^2 = \Omega(1)\Omega(0) = 2R^2$ on the sphere. We also find that $\mathcal{G}_4^{(4)}$ in Figure \ref{phiphi} is finite on the sphere because $\mathcal{A}_3$ is finite. This implies that we do not have a term $(g_1^{n_1}g_2^{n_2})\epsilon^{-1}$ with $n_1 + n_2 = 4$ in the $\eta$ counterterm, which would renormalize the propagator.

Next, we consider the following leading contributions to the propagator from curvature couplings, where the diagrams are depicted in Figure \ref{phiphicurvature}: 
\begin{equation}
\begin{aligned}
  \braket{\sigma_0(x)\sigma_0(y)}&\supset  G_d(x, y)\Bigg(-\eta_{2,0}\mathcal{R}\mathcal{M}_0 +  g_{2,0}\kappa_0\mathcal{R}^2\mathcal{M}_1 - \frac{(Ng_{1,0}^2+g_{2,0}^2)\eta_{2,0}}{2}\left(\mathcal{R}\mathcal{M}_2^{(1)}+\mathcal{R}\mathcal{M}_2^{(3)}\right) \\
  &\qquad \qquad \qquad -(Ng_{1,0}^2\eta_{1,0}+g_{2,0}^2\eta_{2,0})\mathcal{R}\mathcal{M}_2^{(2)}-\frac{Ng_{1,0}g_{2,0}\eta_{1,0}+g^2_{2,0}\eta_{2,0}}{2}\mathcal{R}\mathcal{M}_2^{(4)}\Bigg) ,\\
\end{aligned}
\end{equation}
\begin{equation}
\begin{aligned}
    \braket{\phi^i_0(x)\phi^j_0(y)}&\supset\delta^{ij} G_d(x, y)\Bigg(-\eta_{1,0}\mathcal{R}\mathcal{M}_0 + g_{1,0}\kappa_0\mathcal{R}^2\mathcal{M}_1 - g_{1,0}^2\eta_{1,0}\left(\mathcal{R}\mathcal{M}_2^{(1)}+\mathcal{R}\mathcal{M}_2^{(3)}\right)\\
    &\qquad \qquad \qquad \qquad -g_{1,0}^2(\eta_{1,0}+\eta_{2,0})\mathcal{R}\mathcal{M}_2^{(2)}-\frac{Ng^2_{1,0}\eta_{1,0}+g_{1,0}g_{2,0}\eta_{2,0}}{2}\mathcal{R}\mathcal{M}_2^{(4)} \Bigg) ,
\end{aligned}
\end{equation}
\begin{figure}[t]
\centering
\begin{subfigure}[b]{0.1\textwidth}
\centering
\begin{tikzpicture}[ circ/.style={shape=circle, inner sep=1pt, draw, fill=white, node contents=}] 
\draw (-0.7, 0) to  (0,0)node[cross]{} to  (0.7,0);
\end{tikzpicture} 
\caption{$\mathcal{M}_0$}
\end{subfigure}
\hspace{0.5em}
\begin{subfigure}[b]{0.1\textwidth}
\centering
\begin{tikzpicture}[ circ/.style={shape=circle, inner sep=1pt, draw, fill=white, node contents=}] 
\draw (-0.7, 0) to  (0,0)node[vertex]{} to  (0.7,0);
\draw (0,0) to (0,0.7) node [circ]{};
\end{tikzpicture} 
\caption{$\mathcal{M}_1$}
\end{subfigure}
\hspace{0.5em}
\begin{subfigure}[b]{0.15\textwidth}
\begin{tikzpicture} 
\draw (-0.5, 0) to  (0,0)node[vertex]{} to [out=60,in=120]   (1,0)node[vertex]{} to   (0,0)node[vertex]{} ;
\draw (1, 0) to (1.5, 0)node[cross]{} to (2, 0) ;
\end{tikzpicture} 
\caption{$\mathcal{M}_2^{(1)}$}
\end{subfigure}
\hspace{0.5em}
\begin{subfigure}[b]{0.15\textwidth}
\begin{tikzpicture} 
\draw (-0.5, 0) to  (0,0)node[vertex]{} to (0.5,0)node[cross]{} to 
  (1,0)node[vertex]{} to   (1.5,0) ;
  \draw (0,0) to [out=60,in=120] (1,0);
\end{tikzpicture} 
\caption{$\mathcal{M}_2^{(2)}$}
\end{subfigure}
\begin{subfigure}[b]{0.15\textwidth}
\begin{tikzpicture} 
\draw (-0.5, 0)to  (0,0)node[cross]{} to (0.5,0)node[vertex]{} to 
[out=60,in=120]   (1.5,0)node[vertex]{} to   (0.5,0)node[vertex]{} ;
\draw (2, 0)  to (1.5, 0)node[vertex]{} ;
\end{tikzpicture} 
\caption{$\mathcal{M}_2^{(3)}$}
\end{subfigure}
\hspace{0.5em}
\begin{subfigure}[b]{0.1\textwidth}
\begin{tikzpicture} 
\draw (-0.1, 0) to (0.5,0)node[vertex]{} to 
  (1.1,0) ;
  \draw (0.5,0) to (0.5, 0.3) node[vertex]{} to  [out=160,in=-160] (0.5, 1)  node[cross]{} to  [out=-20,in=20] (0.5, 0.3);
\end{tikzpicture} 
\caption{$\mathcal{M}_2^{(4)}$}
\end{subfigure}
\caption{Leading contributions of curvature terms to the two-point function. The empty dot represents an insertion of $\kappa$ and the cross represents an $\eta$ vertex.}
\label{phiphicurvature}
\end{figure}
where we have pulled out an overall factor of the sphere propagator $ G_d(x, y)$. Evaluating the diagrams, again with the choice $x^2 = 1$ and $y = 0$, we find at the lowest order 
\begin{equation}
    \mathcal{M}_0= R^2(1-\log2+\mathcal{O}(\epsilon)),\quad \mathcal{M}_1=\frac{C_d I_2(\frac{d-2}{2}) }{\text{Vol}(S^d)}R^2(1-\log2+\mathcal{O}(\epsilon)),     
\end{equation}
and for the diagrams quadratic in  $g_{1,0}$ and $g_{2,0}$ we find
\begin{equation}
    \mathcal{M}_2^{(1)}=\mathcal{M}_2^{(3)}=\mathcal{M}_2^{(4)}=R^2\left(-\frac{1-\log2}{3(4\pi)^3\epsilon}+\mathcal{O}(1)\right),\quad\mathcal{M}_2^{(2)}= R^2\left(\frac{1-\log2}{(4\pi)^3\epsilon}+\mathcal{O}(1)\right). 
\end{equation}
Now, imposing that $Z_\phi^{-1}\langle \phi_0^i(x)\phi_0^j(y)\rangle$ and $Z_\sigma^{-1}\langle \sigma_0(x)\sigma_0(y)\rangle$ are finite implies that the $\eta_i$ counterterms in \eqref{eq:ctexpansion} are given by
\begin{equation}
    \delta_{\eta_1}^{(1)}= -\frac{(\eta_1+\eta_2)g_1^2}{(4\pi)^3\epsilon}+\cdots, 
\end{equation}
\begin{equation}
\delta_{\eta_2}^{(1)} = -\frac{N\eta_1g_1^2+\eta_2g_2^2}{(4\pi)^3\epsilon}+\cdots,
\end{equation}
where the dots denote $\mathcal{O}(g_1^{n_1}g_2^{n_2}) $ with $n_1 + n_2 = 6$ as well as higher powers of curvature couplings beginning at $\mathcal{O}(\eta g^4)$.\footnote{Note that the renormalization of $\eta$ is independent of $\kappa$ at leading order. We believe that this can be shown to all orders following arguments similar to those of \cite{Brown:1980qq,Hathrell:1981zb} demanding finiteness of the renormalized stress tensor. Also note that $\delta^{(1)}_{\eta_2}$ can be equivalently obtained from renormalizing the one-point function $\braket{\sigma_0}$.}

\subsubsection{Summary of beta functions for curvature couplings}

The bare couplings in \eqref{eq:baresphere} and \eqref{eq:ctdef} do not change with the arbitrary scale $\mu$, which implies
\begin{equation} \label{eq:BetaKappa}
\beta_\kappa=\frac{\epsilon}{2}\kappa+\frac{Ng_1^2+g_2^2}{12(4\pi)^3} \kappa -\frac{N\eta_1g_1+\eta_2g_2}{30(4\pi)^3} + \cdots,
\end{equation}
where the dots denote $\mathcal{O}(g_1^{n_1}g_2^{n_2}) $ with $n_1 + n_2 = 7$ as well as curvature contributions beginning at $\mathcal{O}(\eta^2g, \eta g^3, \kappa g^4)$, and
\begin{equation}
    \beta_{\eta_1}=-\frac{(2\eta_1+3\eta_2)g_1^2}{3(4\pi)^3}+\cdots, 
\end{equation}
\begin{equation} \label{eq:BetaEta2}
    \beta_{\eta_2}=-\frac{6N\eta_1g_1^2-N\eta_2g_1^2+5\eta_2g_2^2}{6(4\pi)^3}+ \cdots,
\end{equation}
where the dots denote $\mathcal{O}(g_1^{n_1}g_2^{n_2}) $ with $n_1 + n_2 = 6$ as well as higher powers of curvature couplings beginning at $\mathcal{O}(\eta g^4)$. This implies the following fixed points for the curvature couplings: 
\begin{equation} \label{eq:fixedcurv}
    \eta_{1}^*,\eta_{2}^*=\mathcal{O}(\epsilon^2),\quad\kappa^*
    =\mathcal{O}(\epsilon^{\frac{3}{2}}).
\end{equation}
In the next section, we will see that this behavior implies that the curvature couplings do not contribute to the $6-\epsilon$ expansion of $\widetilde{F}$ to order $\epsilon^3$. The renormalization of $b \mathcal{R}^3$ follows from the renormalization of the free energy, which is discussed in the next section. The result for $\beta_b$ is given in \eqref{eq:betab}.

\section{Sphere free energy in dimensional continuation} \label{sec:free}

In this section, we review the computation of the renormalized sphere free energy $F = - \log Z$ and its generalization $\widetilde{F} = - \sin\left(\frac{\pi d}{2}\right)F$ as defined in \cite{Giombi:2014xxa}. In odd dimensions, $\widetilde{F} = (-1)^{\frac{d+1}{2}} F$.  In even dimensions, $\widetilde{F} = \frac{\pi}{2} a $, where $a$ is the generalization of the Weyl anomaly coefficient \cite{Cardy:1988cwa, Komargodski:2011vj} in four dimensions,\footnote{With this definition, $a$ is positive for unitary theories. For example, for a free conformally coupled scalar field, $a=\frac{1}{3}, \frac{1}{90}, \frac{1}{756}$ in $d=2,4,6$, respectively.}
which is related by $c = 3a$ to the central charge studied by \cite{Zamolodchikov:1986gt} in two dimensions. For example, $\widetilde F$ for a conformally coupled free scalar reads on $S^d$ reads
\begin{equation} \label{eq:BosonFreeEnergytilde}
    \widetilde F_{\rm free}=\frac{1}{\Gamma(1+d)}\int_0^1du~u\sin(\pi u)\Gamma\left(\frac{d}{2}+u\right)\Gamma\left(\frac{d}{2}-u\right),
\end{equation}
where we have used \eqref{eq:BosonFreeEnergy}.

The leading term in $\epsilon$-expansion for the sphere free energy of \eqref{cubicO(N)flat} for general $N$ was computed in \cite{Giombi:2014xxa} (also for the case $N=-2$ in \cite{Fei:2015kta}), and the next-to-leading term for large $N$ was computed (without considering curvature vertices) in \cite{Tarnopolsky:2016vvd}. Here we include formulas of the next-to-leading term for general $N$, which we then specify to the cases $N=0,-2,1$, for which we also compute Pad\'{e} approximants in Section \ref{sec:numerics}. Define integrated connected sphere correlation functions
\begin{equation}
    G_n=\int\prod_{i=1}^nd^dx_i\sqrt{g_{x_i}}\braket{\varphi_0^3(x_1)...\varphi_0^3(x_n)}_0^{\text{conn}},
\end{equation}
where $\varphi_0^3\equiv 3 g_{1,0}\sigma_0\phi_0^i\phi_0^i + g_{2,0}\sigma_0^3$. To compute the renormalized sphere free energy to order $\epsilon^2$, we need to consider the renormalization of contributions up to sixth order the couplings $g_{1,0}$ and $g_{2,0}$, for the following reason. The divergent part of these sixth-order terms has a pole in $\epsilon$, which we remove by fixing the counterterms in $b_0$ (which also includes a contribution to remove the one-loop divergence of the conformally coupled scalar \eqref{eq:BosonFreeEnergy}). At the fixed point, $b^*$ then will include a contribution of order $\epsilon^2$ (a factor of $\epsilon^3$ at the fixed point from the sixth order in $g_1$ and $g_2$, multiplied by the $\epsilon^{-1}$ from the $b_0$ counterterms).

In addition to potentially finite contributions from the curvature couplings $\kappa$ and $\eta$ through order $\epsilon^2$ at the fixed point, we must also consider the possibility that they lead to divergences at the same order as $g^6$, which would contribute to renormalization of the free energy. 
Using the fixed points in the previous subsection \eqref{eq:fixedcurv}, the three diagrams
$\kappa^2$  $\begin{tikzpicture} [baseline={([yshift=-.5ex]current bounding box.center)}, circ/.style={shape=circle, inner sep=1pt, draw, fill=white, node contents=}]
\draw (-0.2, 0)node[circ]{} to  (0.2, 0)node[circ]{} ;
\end{tikzpicture}$, $\eta g^2$  $\begin{tikzpicture} [baseline={([yshift=-.5ex]current bounding box.center)}]
\draw[color=black] (0,0) circle [radius=0.2]; 
\draw (0, 0.2)node[vertex]{} to  (0,-0.2)node[vertex]{} ;
\draw (0.2, 0)node[cross]{};
\end{tikzpicture}$, and $\kappa g^3$ $\begin{tikzpicture} [baseline={([yshift=-.5ex]current bounding box.center)}, circ/.style={shape=circle, inner sep=1pt, draw, fill=white, node contents=}]
\draw[color=black] (0,0) circle [radius=0.2]; 
\draw (0, 0.2)node[vertex]{} to  (0,-0.2)node[vertex]{} ;
\draw (0.2, 0)node[vertex]{} to (0.4, 0) node[circ]{};
\end{tikzpicture}$
can potentially contain divergences at the same order as $g^6$ and thus affect the renormalization of $b_0$.
However, the integrated correlation functions with these coefficients are all finite:  the first diagram is proportional to $I_2(\frac{d-2}{2})$ (cf. \eqref{eq:I2}) and the rest are proportional to $\CA_3$ (cf. \eqref{A3final}). 
Therefore, they do not contribute to the renormalized free energy at order $\epsilon^2$.

For the finite terms in the free energy at $\mathcal{O}(\epsilon^2)$ at the fixed point and renormalization at the next order, we can thus simply consider the same terms as in \cite{Tarnopolsky:2016vvd}: 
\begin{equation}\label{eq:FreeEnergy}
\begin{aligned}
    F&=(N+1)F_{\rm free}-\frac{G_2}{2!(3!)^2}-\frac{G_4}{4!(3!)^4}-\frac{G_6}{6!(3!)^6}+ b_0\int d^dx\sqrt{g}\mathcal{R}^3,
\end{aligned}
\end{equation}
where $F_{\rm free}$ is the sphere free energy for conformally coupled free scalar  \eqref{eq:BosonFreeEnergy}. The contribution from $G_n$ and $F_{\rm free}$ for large $N$ was computed in \cite{Tarnopolsky:2016vvd}. Here we write the result for general $N$. A review of previous results and more detailed explanation can be found in Appendix \ref{app:details}.

Expressing the free energy in terms of the renormalized couplings at the fixed point 
\begin{equation}
\beta_{g_1} = \beta_{g_2} = \beta_{\eta_1} = \beta_{\eta_2} = \beta_{\kappa} = \beta_b = 0
\end{equation}
using \eqref{eq:fixedb} and \eqref{eq:fixedcurv}, we obtain $\Tilde{F}=-\sin\frac{\pi d}{2}F$ at $d=6-\epsilon$:
\begin{equation}\label{eq:FtildeResult}
\begin{aligned}
    &\widetilde{F}=(N+1)\widetilde{F}_{\rm free}-\frac{(3Ng_1^{*2}+g_2^{*2})(30+\epsilon(56+15\log(4\pi e^{\gamma_E} \mu^2R^2)))\epsilon}{2^{10}3^45^2(4\pi)^2}\\
    &+\frac{N(26N-148+9(N-8)\log(4\pi e^{\gamma_E} \mu^2R^2))g_1^{*4}\epsilon}{2^{11}3^55(4\pi)^5}-\frac{N(7+3\log(4\pi e^{\gamma_E} \mu^2R^2))g_1^{*3}g_2^*\epsilon}{2^73^45(4\pi)^5}\\
    &+\frac{N(26+9\log(4\pi e^{\gamma_E} \mu^2R^2))g_1^{*2}g_2^{*2}\epsilon}{2^{10}3^55(4\pi)^5}-\frac{(58+27\log(4\pi e^{\gamma_E} \mu^2R^2))g_2^{*4}\epsilon}{2^{11}3^55(4\pi)^5}\\
    &+\frac{N(2(43N+268)g_1^{*6}-12(11N-32)g_1^{*5}g_2^*+(11N+950)g_1^{*4}g_2^{*2}+84g_1^{*3}g_2^{*3}-44g_1^{*2}g_2^{*4})+125g_2^{*6}}{2^{12}3^6 5(4\pi)^8} \\
    &+ \mathcal{O}(\epsilon^4)\,.
\end{aligned}
\end{equation}
This result gives $\widetilde{F}$ at the fixed point for any $N$ at $\mathcal{O}(\epsilon^3)$.  Note we have now explicitly checked that the curvature couplings $\kappa$ and $\eta_1,\eta_2$ do not contribute.

\section{Sphere free energy in long-range approach} \label{sec:NL}

In this section, we study the renormalized sphere free energy in scalar field theories using a $d$-dimensional model with a long-range kinetic term and a slightly relevant perturbation by an operator of dimension $d-\varepsilon$ with $0 < \varepsilon \ll 1$. We will first consider a quartic perturbation in Section \ref{sec:quarticLR} to compare this method with various results in the literature for the Wilson-Fisher fixed point. Long-range models with quartic interactions have a long history \cite{PhysRevLett.29.917, Sak:1973oqx, PhysRevB.15.4344} 
and they have been studied recently in 
\cite{Paulos:2015jfa,Behan:2017dwr,Behan:2017emf,Gubser:2017vgc,Giombi:2019enr, Benedetti:2020rrq,Giombi:2022gjj,Benedetti:2024mqx}.
We will then consider a cubic perturbation in Section \ref{sec:cubicLR}, which corresponds to the long-range versions of the Yang-Lee and related CFTs. 
\subsection{Setup}
Here, we provide an introduction to our approach using the long-range model in the case of a single scalar field on $S^d$ with an action of the general form 
\begin{equation}
  S = S_0 + \lambda_0 \int d^dx \sqrt{g_x}~ O_0(x),
\end{equation}
where $\lambda_0$ is the UV bare coupling, $O_0$ is the bare operator of dimension $d-\varepsilon$, and $S_0$ is the action of a free scalar field with conformally invariant long-range kinetic term \cite{PhysRevLett.29.917,Gubser:2017vgc,Giombi:2019enr,Giombi:2022gjj} 
\begin{align}\label{S0LR}
S_0 = \frac{2^{s-1}\Gamma(\frac{d+s}{2})}{\pi^{\frac{d}{2}} ~\Gamma(-\frac{s}{2})}\int d^d x d^d y \sqrt{g_x}\sqrt{g_y}~\frac{\varphi_0(x) \varphi_0(y)}{D(x,y)^{d+s}}~,
\end{align}
where $D(x, y)$ is defined in \eqref{GD}, $d$ is a fixed integer dimension, and $s$ a parameter that we choose to depend on $\varepsilon$ such that the perturbing (quartic or cubic) interaction for $\varphi$ has consistent dimension $d-\varepsilon$. Note that with this normalization, the momentum-space propagator is simply $|p|^{-s}$. The free two-point function of $\varphi_0$ on the sphere in this model is 
\begin{equation}
  G_{d,s}(x,y) = \frac{C_{d,s}}{D(x,y)^{d-s}}, \ \ \ \ C_{d,s} = \frac{\Gamma(\frac{d-s}{2})}{\pi^{\frac{d}{2}} 2^s ~\Gamma(\frac{s}{2})} .
\end{equation}
In this long-range model, the scaling dimension of $\varphi$ is fixed to be $\frac{d-s}{2}$ and it is not renormalized along the RG flow.
Note that the free propagator for a usual scalar field with local action corresponds to $s=2$. In the case of the $O(N)$ model with quartic interaction, it is expected that when $s$ is near a certain value $s_*$, there is a crossover from the long-range to short-range fixed points \cite{Sak:1973oqx, PhysRevB.15.4344, Behan:2017dwr,Behan:2017emf}. This value is such that the conformal dimension of $\varphi$ is continuous at the crossover, namely $(d-s_*)/2 = \Delta_{\varphi}^{\rm SR}$.  In the long-range region $s< s_*$,  $\varphi$ has no anomalous dimension and its scaling dimension is fixed as $\frac{d-s}{2}$. At the short-range fixed point, $\varphi$ has an anomalous dimension and its dimension is $\frac{1}{2}(d-2 + 2 \gamma_{\varphi}^{\rm SR})$, which implies $s_* = 2 - 2\gamma_{\varphi}^{\rm SR}$.

In long-range models of fixed integer dimension, we can define a generalized free energy
\begin{equation} \label{eq:LRdef}
  \widetilde{F}^{\rm LR} = \begin{cases} (-1)^{\frac{d+1}{2}} F^{\rm LR}, \ &d \text{ odd}, \\
    \frac{\pi}{2} a^{\rm LR},  \ &d \text{  even},\end{cases}
\end{equation}
where $F^{\rm LR} = - \log Z^{\rm LR}$. In odd dimensions, $F^{\rm LR}$ is finite and independent of $R$ for conformal theories  and in even dimensions, $a^{\rm LR}$ is defined as the coefficent of the logarithmic term in $F^{\rm LR} = (-1)^{\frac{d}{2}} a^{\rm LR} \log R + \cdots$, as in short-range theories. This normalization matches the normalization of $\widetilde{F}$ \cite{Giombi:2014xxa} in integer dimensions (the latter is defined for short-range models in dimensional regularization).

Fixing $s$ in terms of $\varepsilon$ by setting the UV dimension of the perturbing operator to be $\Delta_O = d-\varepsilon$, we can compute the generalized free energy in the long-range model
\begin{equation} \label{eq:deltaFtildeLR}
  \widetilde{F}^{\rm LR} = \widetilde{F}_{\rm free}^{\rm LR} + \delta \widetilde{F}^{\rm LR},
\end{equation}
where $\delta \widetilde{F}^{\rm LR}$ is the change in the generalized sphere free energy due to the interaction $O_0$ using perturbation theory in $\varepsilon$,  and $\widetilde{F}_{\rm free}^{\rm LR}$ is the generalized free energy for a free scalar in the long-range model \eqref{S0LR}, defined in odd and even dimensions in analogy with \eqref{eq:LRdef}. Computing $\widetilde{F}_{\rm free}^{\rm LR}$ is equivalent to evaluating the logarithm of the functional determinant of $D(x, y)^{-d-s}$, which is a conformal two-point function of scalar primary operators with scaling dimension $\Delta=\frac{d+s}{2}$ on $S^d$. Such functional determinants correspond to double trace deformations in large $N$ CFT that holographically induce the transition from standard quantization to alternate quantization of the dual bulk operator in AdS \cite{Gubser:2002vv}.
 Both CFT and AdS calculations \cite{Gubser:2002vv, Diaz:2007an, Giombi:2013yva, Giombi:2014xxa, Sun:2020ame, Fraser-Taliente:2024lea} give 
\begin{equation} \label{eq:BosonFreeEnergyLR}
    \widetilde{F}_{\rm free}^{\rm LR}=\frac{1}{\Gamma(1+d)}\int_0^\frac{s}{2}du~u\sin(\pi u)\Gamma\left(\frac{d}{2}+u\right)\Gamma\left(\frac{d}{2}-u\right).
\end{equation}
Taking $s=2$ recovers \eqref{eq:BosonFreeEnergytilde}, corresponding to the usual two-derivative kinetic term.

Generalizing what is believed to happen for the scaling dimensions, we will assume that at the long-range/short-range crossover $s=s_*$, the free energy of the long-range model smoothly goes to that of the short-range one.\footnote{In \cite{Behan:2017dwr}, it was proposed that at the crossover the long-range model goes over to the short-range model plus a decoupled non-local scalar field $\chi$ with $\Delta_{\chi}=(d+s)/2$, rather than to the short-range model alone. If that proposal is understood to hold as a statement about the partition functions of the models, then it would imply that $F^{\rm LR}_{s=s_*}=F^{\rm SR}+F_{\chi}$. This appears to give results for $F^{\rm SR}$ which are far from the ones produced by the 
other methods, such as dimensional continuation. In this paper, we will take the approach that the free energy is continuous at the long-range/short-range crossover.} We will also assume that the same crossover picture discussed in the literature for the quartic models applies to the cubic models as well. Then, to compare with the results of the previous sections and previous literature, in Section \ref{sec:numerics} we evaluate $\widetilde F^{\rm LR}$ at the short-range fixed point $\varepsilon_*$ set by $s_*$. We denote the result $\widetilde{F}^{\rm LRA} \equiv \widetilde{F}^{\rm LR}\big|_{\varepsilon_*}$ for the short-range generalized free energy using the long-range approach. 
   
\subsection{Quartic $O(N)$ model} \label{sec:quarticLR}

Our quartic $O(N)$ model with long-range kinetic term is given by
\begin{equation}
  S = \frac{2^{s-1}\Gamma(\frac{d+s}{2})}{\pi^{\frac{d}{2}} ~\Gamma(-\frac{s}{2})}\int d^d x d^d y \sqrt{g_x}\sqrt{g_y}~\frac{\phi^i_0(x) \phi^i_0(y)}{D(x,y)^{d+s}} + \frac{\lambda_0}{4} \int d^dx \sqrt{g_x}~ \left(\phi_0^i(x) \phi^i_0(x)\right)^2,
\end{equation}
where $i = 1, \ldots, N$ and we take $s = \frac{d+\varepsilon}{2}$ so that $\phi^i_0$ has dimension $\frac{d-\varepsilon}{4}$ and $\left(\phi_0^i \phi^i_0\right)^2$ is a nearly marginal operator of dimension $d-\varepsilon$. Define the renormalized coupling $\lambda$ by 
\begin{equation} \label{eq:lambdaCT}
  \lambda_0 = \mu^{\varepsilon}Z_{\lambda}\lambda, \ \ \ Z_{\lambda} = 1 + \frac{\delta_{\lambda}}{\lambda}. 
\end{equation}
Note that there is no wavefunction renormalization of $\phi^i$ in the long-range model. The beta function for $\lambda$ was computed in \cite{Giombi:2019enr} to order $\lambda^3$:
\begin{equation} \label{eq:betaLambda}
  \beta(\lambda) = - \varepsilon \lambda + \frac{2(N+8)}{(4\pi)^{\frac{d}{2}}\Gamma(\frac{d}{2})}\lambda^2 + \frac{8(5N + 22)}{(4\pi)^d \Gamma(\frac{d}{2})^2}\left(\gamma_E + 2 \psi\left(\frac{d}{4}\right) - \psi\left(\frac{d}{2}\right) \right) \lambda^3  + \mathcal{O}(\lambda^4),
\end{equation}
where $\psi$ is the digamma function. The sphere free energy in the quartic perturbed theory involves the same diagrams as in Figure 1 of \cite{Fei:2015oha}. Define
\begin{equation}
  H_n = \int \prod_{i=1}^{n} d^dx_i \sqrt{g_{x_i}} \langle \phi_0^4(x_1) \cdots \phi_0^4(x_n)\rangle_0^{\text{conn}}. 
\end{equation}
Then we compute the free energy via 
\begin{equation} \label{eq:FNLquartic}
  \delta F^{\rm LR} = - \frac{\lambda_0^2}{2!~4^2} H_2 + \frac{\lambda_0^3}{3!~4^3} H_3 - \frac{\lambda_0^4}{4!~4^4} H_4,
\end{equation}
where the $H_n$ have the same combinatorial prefactors as in \cite{Fei:2015oha}, but the diagrams have different singularities in $\varepsilon$.  Details about the calculation of the free energy are provided in Appendix \ref{app:NLquartic}.\footnote{Note that in the quartic long-range model in two and three dimensions, the only possible curvature counterterm that can appear is $\mathcal{R}$ in two dimensions. We find a $\varepsilon^{-1}$ divergence of $F^{\rm LR}$ beginning at order $\lambda^3$ in two dimensions, which could be renormalized by the $\mathcal{R}$ counterterm.  However, because we work strictly in two dimensions and the divergences do not enter the coefficient of $\log R$, this renormalization does not affect $c$.} In three dimensions, we find at the fixed point
\begin{equation} \label{eq:tildeFNLquartic3D}
  \widetilde{F}^{\rm LR} = N \widetilde{F}_{\rm free}^{\rm LR} - \frac{N(N+2)}{(N+8)^2}\frac{\pi^2}{576}\left(\varepsilon^3 + \frac{3(5N+22)}{(N+8)^2}(2-\pi + 4 \log 2)\varepsilon^4\right) + \mathcal{O}(\varepsilon^5). 
\end{equation}
In two dimensions, the central charge is generally related to $a$-anomaly by $c = 3 a$, as discussed for the long-range case below \eqref{eq:LRdef}, and at the fixed point we find 
\begin{equation} \label{eq:cNLquartic}
  c^{\rm LR} = \frac{6}{\pi} N \widetilde{F}_{\rm free}^{\rm LR} - \frac{N(N+2)}{8(N+8)^2}\left(\varepsilon^3 + \frac{12(5N+22)\log 2}{(8+N)^2} \varepsilon^4 \right) + \mathcal{O}(\varepsilon^5). 
\end{equation}

\subsection{Cubic $O(N)$ model} \label{sec:cubicLR}

We consider the cubic theory with a long-range kinetic term
\begin{equation}
    S = \frac{2^{s-1}\Gamma(\frac{d+s}{2})}{\pi^{\frac{d}{2}} ~\Gamma(-\frac{s}{2})}\int d^d x d^d y \sqrt{g_x}\sqrt{g_y}~\frac{\phi^i_0(x) \phi^i_0(y)+\sigma_0(x) \sigma_0(y) }{D(x,y)^{d+s}} +  \int d^dx \sqrt{g_x}\left( \frac{g_{1,0}}{2} \sigma_0\phi_0^i \phi^i_0 + \frac{g_{2,0}}{6} \sigma_0^3 \right),
\end{equation}
where $i = 1, \ldots, N$ and we take $s = \frac{d+2\varepsilon}{3}$ so that $\phi^i_0$ and $\sigma_0$ have dimension $\frac{d-\varepsilon}{3}$ and the cubic perturbations are nearly marginal operators of dimension $d-\varepsilon$. Renormalized couplings are defined by
\begin{equation}
  g_{i,0} = \mu^{\varepsilon} Z_{g_{i}}g_i, \ \ \ Z_{g_i} = 1 + \frac{\delta_{g_i}}{g_i},\ \ \ i =1,2.
\end{equation}
Note again that the scalar fields themselves are not renormalized in the long-range model. This fact makes the calculation of $\delta_{g_i}$ from corrections to the three-point vertex straightforward, and we include the details in Appendix \ref{app:NLcubic}. The resulting beta functions are  
\begin{equation} \label{eq:betagNL}
  \beta_1=-\varepsilon g_1 - \frac{2 g_1^2(g_1+g_2)}{(4\pi)^{\frac{d}{2}}\Gamma(\frac{d}{2})},\quad \beta_2=-\varepsilon g_2 - \frac{2(N g_1^3+g_2^3)}{(4\pi)^{\frac{d}{2}}\Gamma(\frac{d}{2})}.
\end{equation}
To compute the free energy to order $\varepsilon^2$, we consider the same diagrams up to four points as in \eqref{eq:FreeEnergy}:
\begin{equation} \label{eq:FreeEnergyNLcubic}
\begin{aligned}
  \delta F^{\rm LR}&=-\frac{G_2}{2!(3!)^2}-\frac{G_4}{4!(3!)^4},
\end{aligned}
\end{equation}
where the combinatorial factors of each term are the same as in dimensional regularization, but the integrals have different singularities in the long-range model. They are evaluated for various dimensions in Appendix \ref{app:NLcubic}. In the long-range cubic model, there are possible curvature counterterms in even dimensions, as well as a possible $\sigma \mathcal{R}$ vertex in three dimensions. As noted in the previous subsection, the curvature counterterms in even dimensions do not affect the $\log R$ terms in the free energy, and we do not need to consider them to discuss $a$ and $c$.  In three dimensions, we check that the $\sigma$ one-point function does not have any divergences to order $\mathcal{O}(g_i^3)$ (the diagram $\mathcal{A}_3$ in Figure \ref{phi5g} is finite) and therefore does not contribute to the free energy at order $\varepsilon^2$. At this order, we find at the fixed point
\begin{equation} \label{eq:deltaFLRgen}
   \begin{aligned}
       \delta \widetilde{F}^{\rm LR} = \begin{cases} - \frac{1}{18(4\pi)^2}(3Ng_1^{*2} + g_2^{*2})\varepsilon - \frac{2}{9(4\pi)^4}(3Ng_1^{*4}+ 4 N g_1^{*3} g_2^* + g_2^{*4}) + \mathcal{O}(\varepsilon^3) ,&d=3, \\
       -\frac{\Gamma(\frac{7}{6})^3}{36(4\pi)^{5/2}}(3Ng_1^{*2} + g_2^{*2})\varepsilon - \frac{\Gamma(\frac{7}{6})^3}{36(4\pi)^{9/2}}(3Ng_1^{*4}+ 4 N g_1^{*3} g_2^* + g_2^{*4}) + \mathcal{O}(\varepsilon^3) ,&d=4, \\
       -\frac{\Gamma(\frac{4}{3})^3}{90(4 \pi)^3}(3Ng_1^{*2} + g_2^{*2})\varepsilon - \frac{4\Gamma(\frac{4}{3})^3}{135(4\pi)^6}(3Ng_1^{*4}+ 4 N g_1^{*3} g_2^* + g_2^{*4}) + \mathcal{O}(\varepsilon^3) ,&d=5. \\ \end{cases}
      \end{aligned}
\end{equation}

\section{Numerical results for the sphere free energy} \label{sec:numerics}

In this section, we compare numerical results for the free energy using dimensional continuation and the long-range approach.  We consider quartic theories for small $N$ in the first subsection, and various choices of $N$ in cubic theories in the following subsections.

\subsection{Quartic model}

Here, we compute $\widetilde{F}^{\rm LRA}$, the free energy at the fixed point in the long-range model corresponding to the $d=3$ short-range result by taking $s_* = 3 - 2\Delta_{\phi}^{\rm SR}$ and $\varepsilon_* = 2 s_* -3$ using bootstrap results  \cite{Henriksson:2022rnm} for $\Delta_{\phi}^{\rm SR}$. For $N=1$, $\Delta_{\phi}^{\rm SR} = 0.5181$, for $N=2$, $\Delta_{\phi}^{\rm SR} = 0.5191$, and for $N=3$, $\Delta_{\phi}^{\rm SR} = 0.5189$. The corresponding value $\widetilde{F}^{\rm LRA}$ is presented in the third line of Table \ref{tab:F3dLR}. Here and in subsequent subsections, we find that using the exact value $\widetilde{F}_{\rm free}^{\rm LRA}$ for the free field contribution gives the most sensible results.\footnote{By exact value, we mean substituting $s_*$ into \eqref{eq:BosonFreeEnergyLR} and numerically evaluating the integral. In contrast, we can obtain an $\varepsilon$-expansion of $\widetilde{F}_{\rm free}^{\rm LR}$ using $\varepsilon = 2s -3$. For example, keeping terms up to $\varepsilon^3$, we have in 3$d$
\begin{align*}
\widetilde{F}_{\rm free}^{\rm LR}= 0.04234 +0.03068 \varepsilon-0.0005747 \varepsilon^2-0.005878 \varepsilon^3+\CO\left(\varepsilon^4\right)~.
\end{align*}
Combining this with the leading order correction $\delta\widetilde F^{\rm LR}=-0.00063462 \varepsilon^3+\CO\left(\varepsilon^4\right)$ yields $\widetilde{F}^{\rm LRA}\approx 0.0651$ for $N=1$, which is bigger than $\widetilde F_{\rm free}$. This $\widetilde{F}^{\rm LRA}$ is not a good approximation for Ising model since it violates the $F$-theorem. On the other hand, the $F$-theorem is automatically satisfied if we use the exact value of $\widetilde{F}_{\rm free}^{\rm LRA}$.} We then add to that the contribution of the interactions $\delta\tilde F^{\rm LR}$, evaluated at $\varepsilon=\varepsilon_*$ (we do not use Pad\'{e} approximants in this case as they appear to have poles). We also compare the result of the long-range approach to the fuzzy sphere \cite{Hu:2024pen} and the $\epsilon$-expansion \cite{Fei:2015oha} in Table \ref{tab:F3dLR}. Note that in the fourth line, we use the result of the two-sided Pad\'{e} approximants in Table 1 of \cite{Fei:2015oha} for $\widetilde{F}_{\text{Ising}}^{\text{Pad\'{e}}}$.

\begin{table}[h!]
    \centering
     \renewcommand{\arraystretch}{1.2}
    \begin{tabular}{|c|c|c|c|c|c|}
        \hline
        \text{Quantity} & $N=1$ & $N=2$ & $N=3$   \\\hline
        $N \widetilde{F}_{\rm free}$ &  $0.06381$ & $0.1276$  & $0.1914$   \\\hline
        $N \widetilde{F}_{\rm free}^{\rm LRA}$ & $0.06361$  & $0.1272$  & $0.1908$  \\\hline
        $\widetilde{F}^{\rm LRA}$ & $0.06234$ & $0.1245$ & $0.1868$   \\\hline
        $\widetilde{F}_{\text{Ising}}^{\text{Pad\'{e}}}$ \cite{Fei:2015oha} & $0.06223$  & $0.1243$ & $0.1866$  \\\hline
        $\widetilde{F}$ \text{fuzzy sphere} \cite{Hu:2024pen}& $0.0612$ & -- & --    \\\hline
    \end{tabular}
    \caption{$\widetilde{F}$ in the $d=3$ quartic $O(N)$ model for small $N$. In three dimensions $\widetilde{F} = F$. The exact value of $\widetilde{F}_{\rm free} = \frac{\log 2}{8}- \frac{3\zeta(3)}{16\pi^2}$ in the first line is  from \eqref{eq:BosonFreeEnergytilde}, the second line is the exact value $\widetilde{F}_{\rm free}^{\rm LRA}$  from \eqref{eq:BosonFreeEnergyLR} at $s = s_*$, the third line is the result of the long-range approach \eqref{eq:tildeFNLquartic3D}, and the fourth and fifth lines are $\epsilon$-expansion results from \cite{Fei:2015oha} and fuzzy sphere results from \cite{Hu:2024pen}.  }
    \label{tab:F3dLR}
\end{table}

\begin{table}[h!]
    \centering
     \renewcommand{\arraystretch}{1.1}
    \begin{tabular}{|c|c|c|c|c|c|}
        \hline
        \text{Quantity} & $N=1$ & $N=2$    \\\hline
        $N c_{\rm free}$ & 1 & 2  \\\hline
        $N c_{\rm free}^{\rm LRA}$ & $0.66992$ & $1.3398$  \\\hline
        $c^{\rm LRA}$ & $0.58931$ & $1.1714$     \\\hline
        $c$ (exact) & $\frac{1}{2}$ & 1    \\\hline
    \end{tabular}
    \caption{$c$ in the quartic $O(N)$ model for $N=1,2$. In two dimensions we use  $c = 3 a$. The first line is the exact free value using \eqref{eq:BosonFreeEnergytilde}, the second line uses the exact value $\widetilde{F}_{\rm free}^{\rm LRA}$  from \eqref{eq:BosonFreeEnergyLR} at $s = s_*$, the third line is the result of the long-range approach \eqref{eq:cNLquartic}, and the fourth line is the exact analytic result.}
    \label{tab:F2dLR}
\end{table}

\begin{figure}[h!]
  \centering
  \includegraphics[width=14cm]{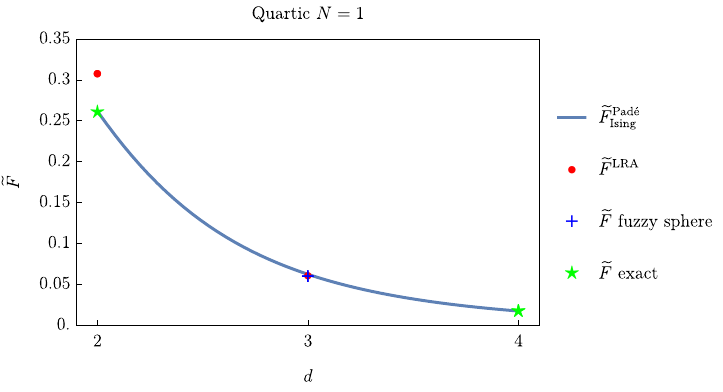}
  \caption{Free energy in the $N=1$ quartic theory between two and four dimensions.  The blue line is the $\epsilon$-expansion result $\widetilde{F}^{\text{Pad\'e}}_{\text{Ising}}$ \cite{Fei:2015oha} based on a two-sided Pad\'{e} approximation, the red dots are the results of the long-range approach $\widetilde{F}^{\rm LRA}$, the blue cross is the fuzzy sphere result, and the green stars are the exact results in two $(\frac{\pi}{12})$ and four $(\frac{\pi}{180})$ dimensions. }
  \label{fig:QuarticLR}
\end{figure}

\begin{figure}[h!]
  \centering
  \includegraphics[width=14cm]{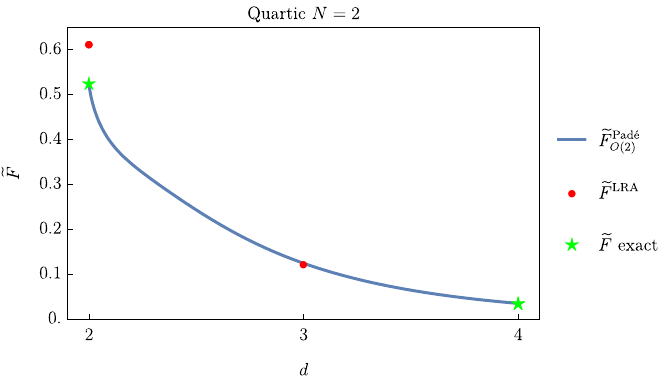}
  \caption{Free energy in the $N=2$ quartic theory between two and four dimensions.  The solid blue line is the $\epsilon$-expansion result $\widetilde{F}_{O(2)}^{\text{Pad\'e}}$ \cite{Fei:2015oha} based on a two-sided Pad\'{e} approximation, the red dots are the results of the long-range approach $\widetilde{F}^{\rm LRA}$, and the green stars are the exact results in two $(\frac{\pi}{6})$ and four $(\frac{\pi}{90})$ dimensions.}
  \label{fig:QuarticLRN2}
\end{figure}

In two dimensions, we can evaluate \eqref{eq:cNLquartic} at $\varepsilon_*$ corresponding to the exact value of $\Delta_{\phi}^{\rm SR}=\frac{1}{8}$ for $N=1,2$ \cite{Nienhuis:1982fx}, using $s_* = 2 - 2\Delta_{\phi}^{\rm SR}$ and $\varepsilon_* = 2 s_* -2$, and compare this to the exact result in Table \ref{tab:F2dLR}. Figures \ref{fig:QuarticLR} and \ref{fig:QuarticLRN2} show results for $\widetilde{F}$ in the $N=1$ and $N=2$ quartic theories, respectively, between two and four dimensions. In Figure \ref{fig:QuarticLR}, we present the $[3,3]$ Pad\'{e} approximant from \cite{Fei:2015oha} as $\widetilde{F}^{\text{Pad\'e}}_{\text{Ising}}$. In Figure \ref{fig:QuarticLRN2}, following \cite{Fei:2015oha}, we present the $[4,2]$ Pad\'{e} approximant of the expanded $f(\epsilon) = \widetilde{F}_{\rm free}(4-\epsilon) + \delta \widetilde{F}$, then add one copy of the exact value $\widetilde{F}_{\rm free}$ to obtain $\widetilde{F}_{O(2)}^{\text{Pad\'e}}$.

\subsection{Cubic Yang-Lee model ($N = 0$)} \label{sec:N0}

The $N=0$ case of the cubic $O(N)$ model, which has applications to the Yang-Lee edge singularity and percolation theory, was considered (aside from studies of general $N$) in flat space by \cite{MACFARLANE197491,PhysRevLett.40.1610,deAlcantaraBonfim:1980pe,OFdeAlcantaraBonfirm_1981,Barbosa:1986kv,Gracey:2015tta,Borinsky:2021jdb} and on the sphere by \cite{Toms:1982af,JACK1986139,PhysRevD.33.2875,PhysRevD.33.2882,Grinstein:2015ina,Osborn:2015rna,Stergiou:2016uqq}. See also \cite{Cardy:2023lha} for a recent review. In $d=2$ the Yang-Lee model corresponds to the $M(2,5)$ minimal model with central charge $c(2,5)=-\frac{22}{5}$ \cite{PhysRevLett.54.1354} and corresponding free energy $\widetilde{F}_{\text{YL}}=\frac{\pi}{6}c(2,5)$. Using the expansion \eqref{eq:FtildeResult} with $N=0$ near $d = 6$, and the exact value at $d = 2$, we have:
\begin{equation}
\begin{aligned}
    \widetilde{F}_{\text{YL}}(d)&=\begin{cases}
        -\frac{11\pi}{15},
        &d=2,\\
        \widetilde{F}_{\rm free}(6-\epsilon)+\frac{\pi\epsilon^2}{25920}+\frac{397\pi\epsilon^3}{7873200}+\mathcal{O}(\epsilon^4), &d=6-\epsilon,
    \end{cases}
    \end{aligned}   
\end{equation}
where $\widetilde{F}_{\rm free}(6-\epsilon)$ denotes the expansion of $\widetilde{F}_{\rm free}$ to order $\epsilon^3$ \eqref{eq:FtildeExpansion}. We can construct a Pad\'{e} approximant to the generalized free energy near $d=6$:
\begin{equation}
\widetilde{F}^{[m,n]}(\epsilon) = \frac{A_0 + A_1 \epsilon + \cdots + A_m \epsilon^m}{1 + B_1 \epsilon + \cdots + B_n \epsilon^n},
\end{equation}
where $m+n \leq 3$, which agrees with the known terms to order $\epsilon^3$ but is expected to provide a better approximation of $\widetilde{F}$. In principle, the approximation can be improved by imposing the known boundary condition at $d=2$. However, we find that doing so leads to approximants with poles in the denominator.  Still, applying the one-point Pad\'{e} approximant to $\widetilde{F}_{\text{YL}}$ at $d=6$, we find that the [1,2] Pad\'{e} has no poles:
\begin{equation} \label{eq:YLPade}
    \widetilde{F}^{[1,2]}_{\text{YL}}=\frac{0.00207777+0.000500993\epsilon}{1-0.742084\epsilon+0.159126\epsilon^2}.
\end{equation}
Values for various choices of $\epsilon$ (choices of $d$) are presented in the last line of Table \ref{tab:N0LR}. It is difficult to extrapolate to $d=2$ with the $\epsilon$ expansion, and we do not include it in the numerics.

\begin{figure}[h!]
  \centering
  \includegraphics[width=14cm]{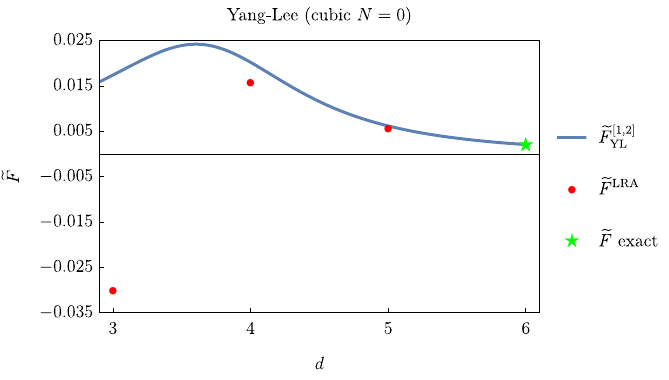}
  \caption{Free energy numerics in $N=0$ cubic Yang-Lee theory in various dimensions.  The solid blue line is the $\epsilon$-expansion result $\widetilde{F}^{[1,2]}_{\text{YL}}$ and the red dots are the results of the long-range approach $\widetilde{F}^{\rm LRA}$. The green star is the exact value $(\frac{\pi}{1512})$ in six dimensions \eqref{eq:FtildeExpansion}. }
  \label{fig:N0LR}
\end{figure}

In the long-range cubic model with $N=0$, using \eqref{eq:deltaFLRgen}, we find the following free energy to order $\varepsilon^2$ at the fixed point:
\begin{equation} \label{eq:FreeEnergyNLcubicN0}
  \begin{aligned}
   & \widetilde{F}^{\rm LR} = \begin{cases} \widetilde{F}_{\rm free}^{\rm LR} + \frac{\varepsilon^2}{288} + \mathcal{O}(\varepsilon^3) , \qquad &d=3, \\
    \widetilde{F}_{\rm free}^{\rm LR} + \frac{\Gamma(\frac{7}{6})^3}{288 \pi^{1/2}}\varepsilon^2 + \mathcal{O}(\varepsilon^3), \qquad &d=4,\\
    \widetilde{F}_{\rm free}^{\rm LR} + \frac{\Gamma(\frac{4}{3})^3}{960}\varepsilon^2+ \mathcal{O}(\varepsilon^3) , \qquad &d=5. \\ \end{cases}
  \end{aligned}
\end{equation}
We can evaluate the long-range free energy at the value $\varepsilon_* = \frac{3 s_* - d}{2}$ with $s_* = d-2 \Delta_{\sigma}^{\rm SR}$ with $\Delta_{\sigma}^{\rm SR}$ at the short-range crossover. The scaling dimension of $\sigma$ in the Yang-Lee CFT has been estimated for $2<d<6$ in a variety of ways, including Pad\' e extrapolations of the $6-\epsilon$ expansion
\cite{PhysRevLett.40.1610,Kompaniets:2021hwg,Borinsky:2021jdb,ArguelloCruz:2025zuq}, high-temperature expansions \cite{Butera:2012tq}, conformal bootstrap \cite{Gliozzi:2013ysa,Gliozzi:2014jsa}, and quantum criticality on regularized spheres \cite{ArguelloCruz:2025zuq,Fan:2025bhc,EliasMiro:2025msj}. 
In $d=3$, the fuzzy sphere result $\Delta_\sigma\approx 0.214$ is the same as the high-temperature expansion result, so we will use this value.
In $d=4$ and $5$ we use the estimates $0.827$ and $1.425$, respectively, which were obtained in \cite{ArguelloCruz:2025zuq} using two-sided Pad\' e extrapolations.
We compare the results of the long-range approach to the $\epsilon$ expansion in Table \ref{tab:N0LR} and Figure \ref{fig:N0LR}.

In Table \ref{tab:N0LR}, $\widetilde{F}$ denotes $-F$ in five dimensions, $\frac{\pi}{2}a$ in four dimensions, and $F$ in three dimensions. The first line is the exact value of $\widetilde{F}_{\rm free}$ in each dimension from \eqref{eq:BosonFreeEnergytilde}, the second line is the exact value $\widetilde{F}_{\rm free}^{\rm LRA}$ in each dimension from \eqref{eq:BosonFreeEnergyLR}  at $s = s_*$, the third line is the result of the long-range approach \eqref{eq:FreeEnergyNLcubicN0} and the last line is the Pad\'{e} approximant in the $\epsilon$-expansion \eqref{eq:YLPade}. Note that the exact value of the central charge is $c = -\frac{22}{5}$ in $d=2$. In the long-range approach in three dimensions,  $\widetilde{F}^{\rm LR}$ is slightly negative, lower than the corresponding $6-\epsilon$ expansion result.

\begin{table}[h!]
    \centering
     \renewcommand{\arraystretch}{1.2}
    \begin{tabular}{|c|c|c|c|}
        \hline
        \text{Dimension}  & $d=5$ & $d=4$ & $d=3$  \\\hline
        $\widetilde{F}_{\rm free}$ & $0.0057430$ & $0.017453$ & $0.063807$ \\\hline
        $\widetilde{F}^{\rm LRA}_{\rm free}$  & $0.0055034$ & $0.012285$ & $-0.049143$  \\\hline
        $\widetilde{F}^{\rm LRA}$   & $0.0058933$ & $0.015894$ & $-0.029837$\\\hline
        $\widetilde{F}^{[1,2]}_{\text{YL}}$  & $0.006183$ & $0.020217$ & $0.017393$  \\\hline
    \end{tabular}
    \caption{Free energy in $N=0$ cubic Yang-Lee theory. $\widetilde{F}$ denotes $-F$ in five dimensions, $\frac{\pi}{2}a$ in four dimensions, and $F$ in three dimensions. The first line is the exact value of $\widetilde{F}_{\rm free}$ in each dimension from \eqref{eq:BosonFreeEnergytilde}, the second line is the exact value $\widetilde{F}_{\rm free}^{\rm LRA}$ in each dimension from \eqref{eq:BosonFreeEnergyLR} at $s=s_*$, the third line is the result of the long-range approach \eqref{eq:FreeEnergyNLcubicN0}, and the last line is the Pad\'{e} approximant in the $\epsilon$-expansion \eqref{eq:YLPade}.}
    \label{tab:N0LR}
\end{table}

\subsection{$OSp(1|2)$ model (cubic $N = -2$)}

In this subsection we consider the cubic $OSp(1|2)$ model on a sphere in $d=6-\epsilon$ dimensions. The $Sp(N)$ model in flat space was proposed in \cite{Fei:2015kta}, where the leading term in the $\epsilon$-expansion of the free energy was computed and for $N=2$ it was shown that the IR fixed point possesses an enhanced global symmetry given by the supergroup $OSp(1|2)$. A more general class of critical field theories with $OSp(1|2M)$ symmetry was studied in \cite{Klebanov:2021sos}. The bare action of the $Sp(2)$ model, including curvature couplings on the sphere, is 
\begin{equation}
\begin{aligned}
    S=\int d^dx\sqrt{g}\left(\partial_\mu\theta_0\partial^\mu\bar{\theta}_0+\frac{1}{2}(\partial_{\mu}\sigma_0)^2+\frac{\xi}{2}\mathcal{R}(\sigma_0^2+2\theta_0\bar{\theta}_0)+g_{1,0}\sigma_0\theta_0\bar{\theta}_0+\frac{1}{6}g_{2,0}\sigma_0^3+\right.\\\left.+\eta_{1,0}\mathcal{R}\theta_0\bar{\theta}_0+\frac{\eta_{2,0}}{2}\mathcal{R}\sigma^2_0+\kappa_0\mathcal{R}^2\sigma_0+b_0\mathcal{R}^3\right),
\end{aligned}
\end{equation}
where $\theta$ is a complex anticommuting scalar. For $g_{2,0} = 2g_{1,0}$, $\eta_{1,0}=\eta_{2,0}$, $\kappa_0=0$ this action possesses a fermionic symmetry with a complex anti-commuting scalar parameter $\alpha$ \cite{Fei:2015kta}
\begin{equation}
    \delta\theta=\sigma\alpha,\quad\delta\bar{\theta}=\sigma\bar{\alpha},\quad\delta\sigma=-\alpha\bar{\theta}+\bar{\alpha}\theta
\end{equation}
that enhances  $Sp(2)$ to $OSp(1|2)$, the smallest supergroup. As a consequence of this symmetry, the scaling dimensions of $\sigma$ and $\theta$ are equal, there is a single cubic coupling $g_0$ and quadratic curvature coupling $\eta_0$, and a linear $\kappa_0$ curvature coupling is forbidden. The beta functions for this theory in $d=6-\epsilon$ are related to the beta functions of the cubic $O(N)$ theory \eqref{cubicO(N)} via the replacement $N = -2$ and specializing to $g \equiv g_2 = 2g_1$, $\eta = \eta_1 = \eta_2$, and $\kappa = 0$. Note that $\beta_{\kappa}$ vanishes exactly in this model, which we check holds to the order we compute in \eqref{eq:BetaKappa}.

\begin{figure}[h!]
  \centering
  \includegraphics[width=14cm]{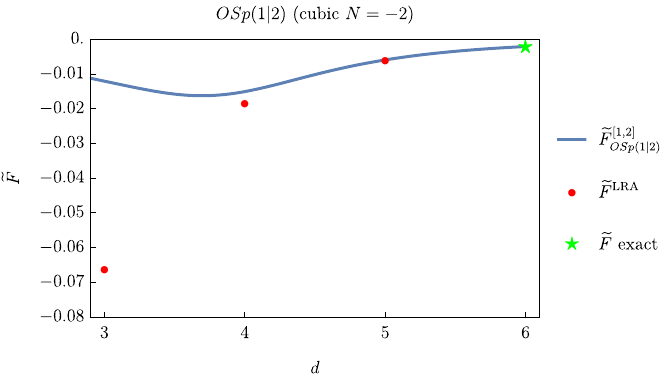}
  \caption{Free energy numerics in $N=-2$ theory $OSp(1|2)$ theory in various dimensions.  The solid blue line is the $\epsilon$-expansion result $\widetilde{F}^{[1,2]}_{OSp(1|2)}$ and the red dots are results of the long-range approach $\widetilde{F}^{\rm LRA}$. The green star is the exact value $(-\frac{\pi}{1512})$ in six dimensions.}
  \label{fig:Nm2LR}
\end{figure}

In two dimensions, the $OSp(1|2)$ model is asymptotically free \cite{Caracciolo:2004hz,2024InMat.237..445B,Bauerschmidt:2019mhu}, which means that the model is conformal in $2<d<6$. 
$\widetilde{F}_{OSp(1|2)}$ approaches  $- \frac{\pi}{3}$, corresponding to $c=-2$, as $d$ approaches $2$ from above.
From the result for general $N$, we find the $6-\epsilon$ expansion of the generalized free energy:
\begin{equation}
\widetilde{F}_{OSp(1|2)}=
-\widetilde{F}_{\rm free}(6-\epsilon)-\frac{\pi\epsilon^2}{43200}-\frac{169\pi\epsilon^3}{5832000}+\mathcal{O}(\epsilon^4)\ .
\end{equation}
Applying the one-sided Pad\'{e} approximant to this expansion, we find that the [1,2] Pad\'{e} has no poles:
\begin{equation} \label{eq:OSpPade}
    \widetilde{F}^{[1,2]}_{OSp(1|2)}=-\frac{0.00207777+0.000539773\epsilon}{1-0.723420 \epsilon+0.164109\epsilon^2}.
\end{equation}
In the long-range cubic model with $N=-2$, using \eqref{eq:deltaFLRgen}, we find the following free energy to order $\varepsilon^2$ at the fixed point:
\begin{equation} \label{eq:FreeEnergyNLcubicNm2}
  \begin{aligned}
    \widetilde{F}^{\rm LR} &= \begin{cases} -\widetilde{F}_{\rm free}^{\rm LR} - \frac{\varepsilon^2}{432} + \mathcal{O}(\varepsilon^3), \qquad &d=3, \\
    -\widetilde{F}_{\rm free}^{\rm LR} - \frac{\Gamma(\frac{7}{6})^3}{432 \pi^{1/2}} \varepsilon^2 + \mathcal{O}(\varepsilon^3), \qquad &d=4,\\
    -\widetilde{F}_{\rm free}^{\rm LR} - \frac{\Gamma(\frac{4}{3})^3}{1440}\varepsilon^2 + \mathcal{O}(\varepsilon^3), \qquad &d=5. \end{cases}
  \end{aligned}
\end{equation}
We can compute the free energy at the values $\varepsilon_*,s_*$ derived from Monte Carlo results in \cite{Deng:2006ur} for $\Delta_{\sigma}^{\rm SR}=\Delta_{\theta}^{\rm SR}$.\footnote{Table I of \cite{Deng:2006ur} gives $\gamma/\nu$ for $q=0$, from which we compute $\Delta_{\sigma}^{\rm SR}=\Delta_{\theta}^{\rm SR} = \frac{d}{2}-\frac{\gamma}{2\nu}$ \cite{Klebanov:2021sos}. In five dimensions $\Delta_{\sigma}^{\rm SR} = 1.46$, in four dimensions $\Delta_{\sigma}^{\rm SR} = 0.920$, and in three dimensions $\Delta_{\sigma}^{\rm SR} = -0.0838$.}  We compare these results to the $6-\epsilon$ expansion in Table \ref{tab:Nm2LR} and Figure \ref{fig:Nm2LR}. The table is set up in parallel with Table \ref{tab:N0LR}.
The two methods give quite close results in $d=4$ and $5$. 
However, since the exact value in two dimensions is $\widetilde{F} = - \frac{\pi}{3} =-1.0472$, it suggests a rapid variation of $\tilde F$ between $2$ and $4$ dimensions. This can make it difficult to estimate $\widetilde{F}$ in $d=3$.

\begin{table}[h!]
    \centering
     \renewcommand{\arraystretch}{1.2}
    \begin{tabular}{|c|c|c|c|}
        \hline
        \text{Dimension}  & $d=5$ & $d=4$     & $d=3$  \\\hline
        $-\widetilde{F}_{\rm free}$  & $-0.0057430$ & $-0.0174531$ & $-0.06381$ \\\hline
        $-\widetilde{F}^{\rm LRA}_{\rm free}$  & $-0.005678$ & $-0.01650$  & $-0.05849$  \\\hline
        $\widetilde{F}^{\rm LRA}$   & $-0.005868$ & $-0.01810$  & $-0.06599$\\\hline
        $\widetilde{F}^{[1,2]}_{OSp(1|2)}$ & $-0.005940$ & $-0.015064$ & $-0.012054$ \\\hline
    \end{tabular}
    \caption{Free energy in  $OSp(1|2)$ model. Here, $\widetilde{F}$ denotes $-F$ in five dimensions, $\frac{\pi}{2}a$ in four dimensions, and $F$ in three dimensions. The first line is the exact value of $\widetilde{F}_{\rm free}$ in each dimension from \eqref{eq:BosonFreeEnergytilde}, the second line is the exact value $\widetilde{F}_{\rm free}^{\rm LRA}$ in each dimension from \eqref{eq:BosonFreeEnergyLR} at $s=s_*$, the third line is the result of long-range approach \eqref{eq:FreeEnergyNLcubicNm2},  and the last line is the Pad\'{e} approximant in the $\epsilon$-expansion \eqref{eq:OSpPade}.}
    \label{tab:Nm2LR}
\end{table}

\subsection{Cubic $N=1$ model} \label{sec:N1}

\begin{figure}[h!]
  \centering
  \includegraphics[width=14cm]{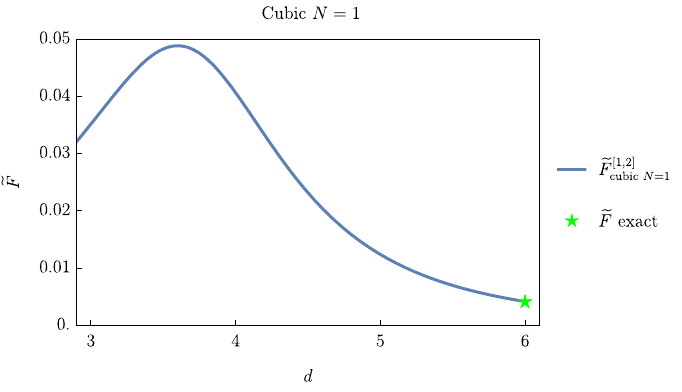}
  \caption{Free energy numerics in $N=1$ cubic theory in various dimensions.  The solid blue line is the $\epsilon$-expansion result $\widetilde{F}^{[1,2]}_{\text{cubic } N=1}$ and the green star is the exact value $\left( \frac{\pi}{756}\right)$ in six dimensions.}
  \label{fig:N1LRcubic}
\end{figure}

The $N=1$ cubic model corresponds in $d=2$ to the $D_5$ modular invariant of $M(3,8)$ minimal model with central charge $c(3,8)=-\frac{21}{4}$, and the corresponding free energy $\widetilde{F}_{\text{cubic }N=1}=\frac{\pi}{6}c(3,8)$  \cite{Fei:2014xta,Klebanov:2022syt,Katsevich:2024jgq}. Using the expansion \eqref{eq:FtildeResult} near $d = 6$ with $N=1$ and the exact value at $d = 2$, we have:\footnote{In the long-range approach, there appears to be no fixed point corresponding to the $M(3, 8)$ universality class. Therefore, this method cannot be used to estimate the sphere free energy.}
\begin{equation}
\begin{aligned}
    \widetilde{F}_{\text{cubic }N=1}(d)&=\begin{cases}
        -\frac{7\pi}{8},
        &d=2,\\
        2\widetilde{F}_{\rm free}(6-\epsilon)+\frac{37\pi\epsilon^2}{479040}+\frac{180905801 \pi\epsilon^3}{1789221585600}+\mathcal{O}(\epsilon^4), &d=6-\epsilon.
    \end{cases}
    \end{aligned}   
\end{equation}
Again, the two-sided Pad\'{e} imposed with the boundary condition at $d=2$ leads to poles in the denominator. Applying the one-sided Pad\'{e} approximant to $\widetilde{F}_{\text{cubic }N=1}$, we find that the [1,2] Pad\'{e} has no poles:
\begin{equation} \label{eq:N1Pade}
    \widetilde{F}^{[1,2]}_{\text{cubic }N=1}=\frac{0.00415555+0.00100299\epsilon}{1-0.741842\epsilon+0.158829\epsilon^2},
\end{equation}
where we present the result in various $d$ in Table \ref{tab:N1} and Figure \ref{fig:N1LRcubic}. 
\begin{table}[h!]
    \centering
     \renewcommand{\arraystretch}{1.2}
    \begin{tabular}{|c|c|c|c|c|c|}
        \hline
        \text{Dimension}  & $d=5$ & $d=4$ & $d=3$  \\\hline
        $2\widetilde{F}_{\rm free}$ & $0.011486$ & $0.034907$ & $0.127614$ \\\hline
        $\widetilde{F}^{[1,2]}_{\text{cubic }N=1}$  & $0.012371$ & $0.040635$ & $0.035131$  \\\hline
    \end{tabular}
    \caption{$6-\epsilon$ expansion for $N=1$ cubic theory. Here, $\widetilde{F}$ denotes $-F$ in five dimensions, $\frac{\pi}{2}a$ in four dimensions, and $F$ in three dimensions. The first line uses the exact value of $\widetilde{F}_{\rm free}$ in each dimension from \eqref{eq:BosonFreeEnergytilde}, and the second line uses the Pad\'{e} approximant \eqref{eq:N1Pade}. }
    \label{tab:N1}
\end{table}

In \cite{Fei:2014xta,Klebanov:2022syt,Katsevich:2024jgq,Delouche:2024tjf}, the RG flow $M(3,10)+\phi_{1,7}\rightarrow M(3,8)$ between the $D$-series modular invariants of two non-unitary minimal models was studied. The $D_6$ modular invariant of $M(3,10)$ describes a pair of the Yang-Lee models $M(2,5)$ \cite{Kausch:1996vq,Quella:2006de,Ardonne_2011}. Let us find the change in $\widetilde{F}$ along the RG flow from $2\widetilde{F}_{\rm YL}$  to $\widetilde{F}_{\text{cubic }N=1}$:
\begin{equation}
\begin{aligned}
   \widetilde{F}_{\text{cubic }N=1} - 2\widetilde{F}_{\rm YL}&=\frac{\pi\epsilon^2}{12934080}+\frac{1018225963 \pi \epsilon^3}{3913027607707200}+\mathcal{O}(\epsilon^4)\\
&=2.42893\cdot10^{-7}\epsilon^2+8.17488\cdot10^{-7}\epsilon^3+\mathcal{O}(\epsilon^4)>0.
\end{aligned}
\end{equation}
Since this change is positive, the generalized $F$-theorem  \cite{Giombi:2014xxa,Fei:2015oha}, which conjectures that the generalized free energy $\widetilde{F}$ decreases for RG flows between unitary theories, is violated in $6-\epsilon$ dimensions for these non-unitary theories. In $d=2$, the $c$ theorem is also violated, since $c(3,10)=- \frac{44}{5}$ 
is smaller than $c(3,8)= -\frac{21}{4}$.
However, the $c_{\rm eff}$ theorem \cite{Castro-Alvaredo:2017udm} is obeyed, since $c_{\rm eff}(3,10)=\frac{4}{5}$ is bigger than $c_{\rm eff}(3,8)=\frac{3}{4}$ \cite{Klebanov:2022syt} (in general, $c_{\rm eff}= c- 24 h_{\rm min}$ as proposed in \cite{Itzykson:1986pk}). This theorem establishes the inequality $c^{\rm UV}_{\rm eff}> c^{\rm IR}_{\rm eff}$ for non-unitary ${\cal PT}$ symmetric 
RG flows \cite{Castro-Alvaredo:2017udm}.  
A natural question, therefore, is whether there exist $d>2$ generalizations of the $c_{\rm eff}$ theorem.

\section*{Acknowledgements}

We are grateful to L. Fei and G. Tarnopolsky for collaboration at very early stages of this project.  We also thank F. De Cesare, J. Kudler-Flam and G. Tarnopolsky for very useful discussions. We additionally thank Ludo Fraser-Taliente and Lorenzo Benfatto for alerting us to some typos in Appendix E that were present in earlier versions of this paper. This work was supported in part by the US National Science Foundation Grant No.~PHY-2209997 and by the Simons Foundation Grant No.~917464.

\appendix

\section{Flat space renormalization} \label{app:flat}

In this appendix, we summarize previous results for the flat-space renormalization of the cubic $O(N)$ model. In minimal subtraction, the counterterms \eqref{eq:ctdef_flat} have the structure 
\begin{equation}
    \delta_{\phi,\sigma}=\sum_{n=1}^\infty\frac{\delta^{(n)}_{\phi,\sigma}(g_1,g_2)}{\epsilon^n},\quad\delta_{g_1,g_2}=\sum_{n=1}^\infty\frac{\delta^{(n)}_{g_1,g_2}(g_1,g_2)}{\epsilon^n}.
\end{equation}
In perturbation theory, powers of $g_1$ and $g_2$ appear as follows:
\begin{equation}
    \delta_{\phi,\sigma}^{(n)}=\sum_k\delta_{\phi,\sigma}^{(n,2k)},\quad\delta_{g_1,g_2}^{(n)}=\sum_k\delta_{g_1,g_2}^{(n,2k+1)},\quad\{\delta_{\phi,\sigma}^{(n,k)},\delta_{g_1,g_2}^{(n,k)}\}\propto g_1^ig_2^j, \ \  i+j = k.
\end{equation}

The one-loop result was first obtained in \cite{Fei:2014yja}, then the three-loop result in \cite{Fei:2014xta}, four-loop result in \cite{Gracey:2015tta} and five-loop result in \cite{Kompaniets:2021hwg}. We use the two-loop counterterms: 
\begin{equation}
    \delta_{\phi}^{(1,2)}=-\frac{g_1^2}{3(4\pi)^3},\quad\delta_{\phi}^{(1,4)}=\frac{g_1^2(g_1^2(11N-26)-48g_1g_2+11g_2^2)}{432(4\pi)^6},
\end{equation}
\begin{equation} \label{eq:sigmact}
    \delta_\sigma^{(1,2)}=-\frac{Ng_1^2+g_2^2}{6(4\pi)^3},\quad\delta_{\sigma}^{(1,4)}=-\frac{2Ng_1^4+48Ng_1^3g_2-11Ng^2_1g^2_2+13g_2^4}{432(4\pi)^6},
\end{equation}
and two-loop wavefunction renormalization 
\begin{align} \label{eq:Zphi}
    Z_\phi&=1-\frac{g_1^2}{3(4\pi)^3\epsilon}-\frac{g_1^2(g_1^2(N-10)-12g_1g_2+g_2^2)}{36(4\pi)^6\epsilon^2}+\frac{g_1^2(g_1^2(11N-26)-48g_1g_2+11g_2^2)}{432(4\pi)^6\epsilon},\\
    Z_\sigma&=1-\frac{Ng_1^2+g_2^2}{6(4\pi)^3\epsilon}+\frac{4Ng_1^4+12Ng_1^3g_2-Ng_1^2g_2^2+5g_2^4}{36(4\pi)^6\epsilon^2}-\frac{2Ng_1^4+48Ng_1^3g_2-11Ng^2_1g^2_2+13g_2^4}{432(4\pi)^6\epsilon}.
    \label{eq:Zsigma}
\end{align}
The beta-functions for the cubic $O(N)$ model were computed in \cite{Fei:2014yja,Fei:2014xta,Gracey:2015tta,Kompaniets:2021hwg}. At two loops,
\begin{align} \label{eq:gbeta}
    \beta_{g_1}&=-\frac{\epsilon}{2}g_1+\frac{g_1((N-8)g_1^2-12g_1g_2+g_2^2)}{12(4\pi)^3}\nonumber\\
    &\qquad -\frac{g_1((86N+536)g_1^4-12(11N-30)g_1^3g_2+(11N+628)g_1^2g_2^2+24g_1g_2^3-13g_2^4)}{432(4\pi)^6},\\
    \beta_{g_2}&=-\frac{\epsilon}{2}g_2-\frac{4Ng_1^3-Ng_1^2g_2+3g_2^3}{4(4\pi)^3} -\frac{24Ng_1^5+322Ng_1^4g_2+60Ng_1^3g_2^2-31Ng_1^2g_2^3+125g_2^5}{144(4\pi)^6}.
\end{align}
We also have the two-loop relation between bare and renormalized couplings
\begin{align} \label{eq:g10}
g_{1,0} &= \mu^{\epsilon/2}\Bigg( g_1 + \frac{g_1((N-8)g_1^2-12g_1g_2+g_2^2)}{12(4\pi)^3 \epsilon}  \nonumber\\
    &\qquad  -\frac{g_1((86N+536)g_1^4-12(11N-30)g_1^3g_2+(11N+628)g_1^2g_2^2+24g_1g_2^3-13g_2^4)}{864(4\pi)^6\epsilon}  \\
    &\qquad  +\frac{g_1((N(80+3N)+352)g_1^4+24(28-5N)g_1^3g_2+10(N+24)g_1^2g_2^2+72g_1g_2^3-17g_2^4)}{288(4\pi)^6\epsilon^2 } \Bigg), \nonumber\\
    g_{2,0} &= \mu^{\epsilon/2}\Bigg(g_2-\frac{4Ng_1^3-Ng_1^2g_2+3g_2^3}{4(4\pi)^3\epsilon} -\frac{24Ng_1^5+322Ng_1^4g_2+60Ng_1^3g_2^2-31Ng_1^2g_2^3+125g_2^5}{288(4\pi)^6\epsilon} \nonumber \\
    &\qquad \qquad  + \frac{24N(4-N) g_1^5 + N(5N+128)g_1^4g_2 + 72N g_1^3 g_2^2 - 34 N g_1^2 g_2^3 + 81 g_2^5}{96 (4\pi)^6 \epsilon^2} \Bigg).
    \label{eq:g20}
\end{align}
For reference, we include the fixed points in $d=6-\epsilon$ for the specific $N$ considered in Section \ref{sec:numerics}:
\begin{equation} \label{eq:cases}
\begin{aligned}
   & N=0: \quad g_2^*=i\sqrt{\frac{2(4\pi)^3\epsilon}{3}}\left(1+\frac{125}{324}\epsilon+\mathcal{O}(\epsilon^2)\right), \quad g_1^* = 0, \\
   & N=-2: \quad g_2^* =2 g_1^*=i\sqrt{\frac{4(4\pi)^3\epsilon}{5}}\left(1+\frac{67}{180}\epsilon+\mathcal{O}(\epsilon^2)\right), \\
    &N=1: \quad \begin{cases}
g_1^*=40i\sqrt{\frac{6\pi^3\epsilon}{499}}\left(1+\frac{2633149}{7470030}\epsilon+\mathcal{O}(\epsilon^2)\right), \\
g_2^*=48i\sqrt{\frac{6\pi^3\epsilon}{499}}\left(1+\frac{227905}{498002}\epsilon+\mathcal{O}(\epsilon^2)\right). \\
    \end{cases}
    \end{aligned}
    \end{equation}
Note that for $N=1$ there is also an unstable fixed point 
\begin{equation}
    g_1^*=g_2^*=i\sqrt{\frac{(4\pi)^3\epsilon }{3}}\left(1+\frac{125\epsilon}{324}+\mathcal{O}(\epsilon^2)\right)\,,
\end{equation}
which is the sum of two $N=0$ theories \cite{Fei:2014xta}. We discuss the flow between them in Section \ref{sec:N1}. 
\section{Integral conventions on the sphere} \label{app:conventions}
In this appendix, we review our conventions for computing Feynman integrals on sphere, following \cite{Fei:2015oha}. We use the conformally flat metric on the $d$-dimensional sphere of radius $R$:
\begin{align}
    ds^2 = \Omega(x)^2 dx^2, \quad \Omega (x) \equiv \frac{2R}{1+x^2},
\end{align}
where $x\in\mathbb{ R}^d$. Its volume is 
\begin{equation} \label{eq:vold}
    {\text{Vol}(S^d)} = \int d^d x \sqrt{g} = \frac{2 \pi^{\frac{d+1}{2}}R^d}{\Gamma\left(\frac{d+1}{2}\right)},
\end{equation}
and its Ricci scalar curvature is 
\begin{equation} \label{eq:Ricci}
\mathcal{R} = \frac{d(d-1)}{R^2}. 
\end{equation}
The propagator $\mathbb{G}_d$ for a massless scalar on $\mathbb{R}^d$ is
\begin{align}\label{eq:flatprop}
    \mathbb G_d (x, y)=  \frac{C_d}{|x- y|^{d-2}}, \quad C_d\equiv \frac{\Gamma(\frac{d}{2}-1)}{4\pi^{\frac{d}{2}}}.
\end{align}
Because the sphere is conformally flat, the propagator $G_d$ of a conformally coupled scalar on $S^d$ is related to the propagator of a massless scalar on $\mathbb R^d$ by the Weyl factor $\Omega$:
\begin{align}\label{GD}
    G_d(x, y) = \frac{C_d}{D(x,y)^{d-2}}, \quad D(x,y) \equiv \sqrt{\Omega(x)\Omega(y)}|x-y|.
\end{align}
Note that the function $D(x,y)$ is $SO(d+1)$ invariant. Two well-known exactly solvable integrals on sphere are \cite{Cardy:1988cwa,Klebanov:2011gs,Drummond:1975yc,Drummond:1977dn}
\begin{align}\label{eq:I2}
    I_2(\Delta)=\int\frac{d^dxd^dy~\Omega^d(x)\Omega^d(y)}{D(x,y)^{2\Delta}}=2^{1+d-2\Delta}\pi^{d+\frac{1}{2}}R^{2(d-\Delta)}\frac{\Gamma(\frac{d}{2}-\Delta)}{\Gamma(\frac{d+1}{2})\Gamma(d-\Delta)},
\end{align}
\begin{align}\label{eq:I3}
    I_3(\Delta)=\int\frac{d^dxd^dyd^dz~\Omega^d(x)\Omega^d(y)\Omega^d(z)}{\left[D(x,y)D(y,z)D(z,x)\right]^{\Delta}}=8\pi^{\frac{3(1+d)}{2}}R^{3(d-\Delta)}\frac{\Gamma(d-\frac{3\Delta}{2})}{\Gamma(\frac{d+1-\Delta}{2})^3\Gamma(d)}.
\end{align}

\section{Details of free energy calculations} \label{app:details}

In this appendix, we include a summary of the terms that contribute to the sphere free energy in the $O(N)$ cubic model at sixth order in $g_{1,0}$ and $g_{2,0}$, computed in \cite{Tarnopolsky:2016vvd}.

The sphere free energy for a conformally coupled free scalar in dimensional regularization on $S^d$ is \cite{Diaz:2007an,Giombi:2014xxa}: 
\begin{equation} \label{eq:BosonFreeEnergy}
    F_{\rm free}=-\frac{1}{\sin\left(\frac{\pi d}{2}\right)\Gamma(1+d)}\int_0^1du~u\sin(\pi u)\Gamma\left(\frac{d}{2}+u\right)\Gamma\left(\frac{d}{2}-u\right).
\end{equation}
Near every even $d$ this expression has a simple pole, with coefficient that reproduces the known $a$-anomalies. In $d=6-\epsilon$, 
\begin{equation} 
    F_{\rm free}=-\frac{1}{756\epsilon} + \mathcal{O}(1).
\end{equation}
The expansion of $\widetilde{F}_{\rm free}(6-\epsilon)$ is
\begin{equation} \label{eq:FtildeExpansion}
  \widetilde{F}_{\rm free}(6-\epsilon) = \frac{\pi}{1512} + 0.002042876 \epsilon + 0.001064155\epsilon^2 + 0.000396195 \epsilon^3+\mathcal{O}(\epsilon^4).
\end{equation}
The free energy \eqref{eq:FreeEnergy} up to order $\epsilon^2$ involves the diagrams 
\begin{align}
    & G_2=3!t_2C_d^3I_2\left(\frac{3(d-2)}{2}\right),\\
    & G_4=3(3!)^3(3t_{41}G_4^{(1)}+2t_{42}G_4^{(2)}),\\
    & G_6=15(3!)^5(18t_{61}G_6^{(1)}+6t_{62}G_6^{(2)}+9t_{63}G_6^{(3)}+36t_{64}G_6^{(4)}+24t_{65}G_6^{(5)}+4t_{66}G_6^{(6)}),
\end{align}
which are shown in Figure 4 of \cite{Tarnopolsky:2016vvd}, where the $t$-coefficients are
\begin{equation}
\begin{aligned} \label{eq:tcoefs}
    & t_2=3Ng_{1,0}^2+g_{2,0}^2,\quad t_{41}=(N+4)Ng_{1,0}^4+2Ng_{1,0}^2g_{2,0}^2+g_{2,0}^4,\\
    & t_{42}=3Ng_{1,0}^4+4Ng_{1,0}^3g_{2,0}+g_{2,0}^4,\\
    & t_{61}=4N(N+1)g_{1,0}^6+N(N+4)g_{1,0}^4g_{2,0}^2+2Ng_{1,0}^2g_{2,0}^4+g_{2,0}^6,\\
    & t_{62}=8Ng_{1,0}^6+(Ng_{1,0}^2+g_{2,0}^2)^3,\quad t_{63}=N^2g_{1,0}^4(2g_{1,0}+g_{2,0})^2+4Ng_{1,0}^3g_{2,0}^3+g_{2,0}^6,\\
    & t_{64}=N(N+4)g_{1,0}^6+2N(N+2)g_{1,0}^5g_{2,0}+Ng_{1,0}^4g_{2,0}^2+2Ng_{1,0}^3g_{2,0}^3+Ng_{1,0}^2g_{2,0}^4+g_{2,0}^6,\\
    & t_{65}=N(N+3)g^6_{1,0}+6Ng^5_{1,0}g_{2,0}+3Ng^4_{1,0}g_{2,0}+2Ng^3_{1,0}g^3_{2,0}+g^6_{2,0},\\
  & t_{66}=6Ng^6_{1,0}+9Ng^4_{1,0}g^2_{2,0}+g_{2,0}^6.
\end{aligned}
\end{equation}
Evaluating the diagrams in $d=6-\epsilon$ gives
\begin{equation}
    G_4^{(k)}=C_d^6(2R)^{12-3d}\text{Vol}(S^d)e^{\frac{3\gamma_E}{2}(d-6)+\frac{\pi^2}{16}(d-6)^2}\pi^{\frac{3d}{2}}\begin{cases}
        \frac{1}{18\epsilon}+\frac{43}{162}+\frac{2857}{3888}\epsilon+\mathcal{O}(\epsilon^2), &k=1,\\
        -\frac{1}{3\epsilon}-\frac{3}{2}-\frac{191}{48}\epsilon+\mathcal{O}(\epsilon^2), &k=2,
    \end{cases}
\end{equation} 
\begin{equation}
    G_6^{(k)}=C_d^9(2R)^{18-4d}\text{Vol}(S^d)e^{\frac{5\gamma_E}{2}(d-6)+\frac{\pi^2}{16}(d-6)^2}\pi^{\frac{5d}{2}}\begin{cases}
        -\frac{1}{54\epsilon^2}-\frac{389}{2592\epsilon}-\frac{87}{128}+\mathcal{O}(\epsilon), &k=1,\\
        -\frac{1}{36\epsilon^2}-\frac{139}{648\epsilon}-\frac{3547}{3888}+\mathcal{O}(\epsilon), &k=2,\\
        \frac{1}{384}+\mathcal{O}(\epsilon), &k=3,\\
        \frac{1}{9\epsilon^2}+\frac{733}{864\epsilon}+\frac{4675}{1296}+\mathcal{O}(\epsilon), &k=4,\\
        -\frac{1}{3\epsilon^2}-\frac{703}{288\epsilon}+\mathcal{O}(1), &k=5,\\
        -\frac{5}{24\epsilon}+\frac{\zeta(3)}{8}-\frac{475}{288}+\mathcal{O}(\epsilon), &k=6.
    \end{cases}
\end{equation}
Note that these expressions include a small correction to \cite{Tarnopolsky:2016vvd}, specifically the $\mathcal{O}(1)$ term in $G_6^{(3)}$ (which does not change the result for the free energy at the order considered in \cite{Tarnopolsky:2016vvd}). We present the calculation of $G_6^{(3)}$ as an example in the next subsection.

To renormalize the free energy to sixth order in $g_1$ and $g_2$, we compute the contributions to $F$ from $G_2$, $G_4$, and $G_6$ and use \eqref{eq:g10} and \eqref{eq:g20} to express these contributions in terms of renormalized couplings. Then, we fix the counterterms in $b_0$ to remove the divergences in the free energy of the conformally coupled scalar \eqref{eq:BosonFreeEnergy} as well as divergences in the terms in $F$ of order $\mathcal{O}(g_1^{n_1}g_2^{n_2})$ with $n_1 + n_2 = 6$ \cite{Tarnopolsky:2016vvd}:
\begin{equation}
\begin{aligned}
    \delta^{(1)}_b=\frac{N+1}{756\cdot450(4\pi)^3}+b_{61} + \cdots,\\
    \end{aligned}
\end{equation}
where dots denote higher powers of $g_1$ and $g_2$ as well as higher contributions of the curvature couplings, and
\begin{equation}
\begin{aligned}
b_{61} = \frac{N(2(43N+268)g_1^6-12(11N-32)g_1^5g_2+(11N+950)g_1^4g_2^2+84g_1^3g_2^3-44g_1^2g_2^4)+125g_2^6}{2^{12}3^85^3(4\pi)^{12}}.
\end{aligned}
\end{equation}
The beta function is then
\begin{equation} \label{eq:betab}
\begin{aligned}
    &\beta_b=\epsilon b+\frac{N+1}{756\cdot450(4\pi)^3}+4b_{61} + \cdots\\
\end{aligned}
\end{equation}
where dots denote higher powers of $g_1$ and $g_2$ as well as higher contributions from curvature couplings, with fixed point
\begin{equation} \label{eq:fixedb}
\begin{aligned}
    &b^*=-\frac{N+1}{756\cdot450(4\pi)^3\epsilon}-\frac{4b_{61}^*}{\epsilon} + \mathcal{O}(\epsilon^3).\\
\end{aligned}
\end{equation}
After renormalization, we find the free energy up to fourth order in $g_1$ and $g_2$, which is given at the fixed point by \eqref{eq:FtildeResult}.

\subsection{Analytic calculation of $G^{(3)}_6$} \label{app:analyticA}
The (one-particle-reducible) diagram $G_6^{(3)}=\begin{tikzpicture} [baseline={([yshift=-.5ex]current bounding box.center)}]
\draw[color=black] (0,0) circle [radius=0.2]; 
\draw[color=black] (0.6,0) circle [radius=0.2]; 
\draw (0.2, 0)node[vertex]{} to (0.4, 0)node[vertex]{} ;
\draw (0, 0.2)node[vertex]{} to  (0,-0.2)node[vertex]{} ;
\draw (0.6, 0.2)node[vertex]{} to  (0.6,-0.2)node[vertex]{} ;
\end{tikzpicture} $ admits a simple analytic evaluation. To show this, we focus on its irreducible component $\begin{tikzpicture} [baseline={([yshift=-.5ex]current bounding box.center)}]
\draw[color=black] (0,0) circle [radius=0.2]; 
\draw (0.2, 0)node[vertex]{} ;
\draw (0, 0.2)node[vertex]{} to  (0,-0.2)node[vertex]{} ;
\end{tikzpicture} $
which is denoted by $\CA_3$ in Figure \ref{phi5g}. It takes the following form:
\begin{equation}
\mathcal{A}_3\equiv\int\prod_{i=1}^2\left(d^dx_i\Omega^d(x_i)G_d(x_i,x_0)\right)G_d(x_1,x_2)^2~.
\end{equation}
This integral is independent of $x_0$ due to the $SO(d+1)$ symmetry. Therefore we can set $x_0=0$ in the stereographic coordinates. After performing the inversion $x_i^\mu\to x_i^\mu/x_i^2$  for $x_1$ and $x_2$, we find 
\begin{align}
    \mathcal{A}_3 &=(2R)^{8-2d}C_d^4\int\prod_{i=1}^2\frac{d^dx_i}{(1+x_i^2)^{3-\frac{d}{2}}}\frac{1}{x^{2(d-2)}_{12}}~.
 \end{align}
 The remaining integral has been studied in Appendix B of \cite{Fei:2015oha}. Following the notation of that paper, the result of $\CA_3$ can be expressed as 
 \begin{align}\label{A3final}
 \mathcal{A}_3=(2R)^{8-2d}C_d^4\,\Gamma_0\left(3-\frac{d}{2},3-\frac{d}{2},d-2\right) = -\frac{1}{8(4\pi)^6R^4}+\CO(\epsilon)~,
\end{align}
where
\begin{equation}
\Gamma_0(a_1,a_2,b)=\frac{\pi^d\Gamma(\frac{d}{2}-b)\Gamma(a_1+b-\frac{d}{2})\Gamma(a_2+b-\frac{d}{2})\Gamma(a_1+a_2+b-d)}{\Gamma(\frac{d}{2})\Gamma(a_1)\Gamma(a_2)\Gamma(a_1+a_2+2b-d)}.
\end{equation}
The result for $\CA_3$ first appeared in \cite{Drummond:1975yc,Drummond:1977dn}.
Then, by gluing two copies of $\CA_3$, we get
\begin{align}
    G_6^{(3)}&=C_d\, \CA_3^2 \,I_2\left(\frac{d-2}{2}\right)=(2R)^{18-3d}C_d^9\Gamma^2_0\left(3-\frac{d}{2},3-\frac{d}{2},d-2\right)\Gamma_0\left(1+\frac{d}{2},1+\frac{d}{2},\frac{d-2}{2}\right)\nonumber\\
    &=(2R)^{18-4d}\text{Vol}(S^d)C_d^9e^{\frac{5\gamma_E}{2}(d-6)+\frac{\pi^2}{16}(d-6)^2}\pi^{\frac{5d}{2}}\left(\frac{1}{384}+\frac{77\epsilon}{4608}+\mathcal{O}(\epsilon^2)\right),
\end{align}
which is finite in the $\epsilon\to 0$ limit.

\section{Beta functions for curvature couplings for the cubic $N=0$ theory} \label{app:comparison}

In this appendix, we summarize our results for the cubic $N=0$ curvature coupling renormalizations and comment on previous literature. Our results for the curvature coupling beta functions at $N=0$ are (with $g_1\rightarrow0,g_2\rightarrow g,\eta_1\rightarrow0,\eta_2\rightarrow\eta$):
\begin{equation} \label{eq:ourcurvebeta}
\begin{aligned}
  \beta_\kappa&=\frac{\epsilon}{2}\kappa+\frac{\kappa g^2}{12(4\pi)^3}-\frac{\eta g}{30(4\pi)^3}+ 0 \cdot g^3 + 0 \cdot g^5 + \cdots , \\
  \beta_\eta&=-\frac{5\eta g^2}{6(4\pi)^3}+  0 \cdot g^4 + \cdots,\\
\end{aligned}
\end{equation}
where the dots in $\beta_{\kappa}$ denote terms at  $\mathcal{O}(g^7)$  as well as curvature contributions beginning at $\mathcal{O}(\eta^2g, \eta g^3, \kappa g^4)$ and the dots in $\beta_{\eta}$ denote terms at $\mathcal{O}(g^6)$ as well as higher powers of curvature couplings beginning at $\mathcal{O}(\eta g^4)$.

Our results for $\beta_{\kappa}$ agree with the one-loop results of \cite{Toms:1982af,JACK1986139,PhysRevD.33.2882} but differ from the two-loop results of \cite{JACK1986139,PhysRevD.33.2882} by the $g^3$ and $g^5$ terms, which they find to be nonzero.\footnote{The computation of $\beta_\kappa$ involves the beta function of $g$. We believe that there is a typo in the two-loop counterterm of $g$ in  \cite{JACK1986139}.} The absence of a $g^3$ correction in our beta function is because the diagram $\CA_3$ is finite, as found in \cite{Drummond:1975yc,Drummond:1977dn}. In fact, \cite{Drummond:1977dn} comments that there should be no such $g^3$ term in $\beta_{\kappa}$. We have reviewed the analytic calculation of $\mathcal{A}_3$ in Appendix \ref{app:analyticA}.

Our results for $\beta_{\eta}$ agree with \cite{Toms:1982af,JACK1986139,PhysRevD.33.2875,PhysRevD.33.2882} at one-loop order. The leading mass anomalous dimension is $\gamma_{\sigma^2}^{\rm irred}(g) = -  \frac{5 g^2}{6 (4\pi)^3}$, where the superscript means that only the one-particle-irreducible diagrams are included. The first term in our result \eqref{eq:ourcurvebeta} for $\beta_\eta$ has the form $\eta \gamma_{\sigma^2}^{\rm irred}(g)$ in agreement with \cite{Toms:1982af,JACK1986139,PhysRevD.33.2875,PhysRevD.33.2882}. However, unlike \cite{JACK1986139,PhysRevD.33.2875,PhysRevD.33.2882}, we do not find the $g^4$ term in $\beta_\eta$, again because the diagram $\mathcal{A}_3$ in Figure \ref{phi5g} is finite. Therefore the two-point function contribution from $\mathcal{G}_4^{(4)}$ in Figure \ref{phiphi} is finite.

Renormalization of curvature terms in the background field method was also considered by \cite{Grinstein:2015ina,Osborn:2015rna,Stergiou:2016uqq}, who found results agreeing with \cite{JACK1986139} for the pure curvature terms, but did not compute two-loop beta functions for the curvature couplings to scalars.

\section{Details of long-range model calculations} \label{app:NL}

In this appendix, we include some details of the renormalized free energy calculations in Section \ref{sec:NL}.

\subsection{Quartic model} \label{app:NLquartic}

Note that the beta function \eqref{eq:betaLambda} implies that in \eqref{eq:lambdaCT} 
\begin{equation} \label{eq:renormLambda}
  \delta_{\lambda} = \frac{2(N+8)}{(4\pi)^{\frac{d}{2}}\Gamma(\frac{d}{2})}\frac{\lambda^2}{\varepsilon} + \frac{4(N+8)^2}{(4\pi)^{d}\Gamma(\frac{d}{2})^2}\frac{\lambda^3}{\varepsilon^2} + \frac{4(5N + 22)}{(4\pi)^d \Gamma(\frac{d}{2})^2}\left(\gamma_E + 2 \psi\left(\frac{d}{4}\right) - \psi\left(\frac{d}{2}\right) \right) \frac{\lambda^3}{\varepsilon}  + \mathcal{O}(\lambda^4). 
\end{equation}
The diagrams that contribute to \eqref{eq:FNLquartic} are shown in Figure 1 of \cite{Fei:2015oha}. In the long-range model, they are
\begin{equation}
  \begin{aligned}
    H_2 &= 8N(N+2)C_{d,s}^4I_2(d-\varepsilon), \\
    H_3 &= 64N(N+2)(N+8)C_{d,s}^6I_3(d-\varepsilon), \\
    H_4 &= 32(4!)N(N+2)C_{d,s}^8\left((N^2+ 6N + 20) H_4^{(1)} + 8(N+2)H_4^{(2)} + 4(5N+22)H_4^{(3)}\right),
  \end{aligned}
\end{equation}
where the $H_4^{(k)}$ are diagrams that are evaluated with Mellin-Barnes integrals directly in various integer dimensions:
\begin{equation}
  \begin{aligned}
  H_{4}^{(1)} &= (2R)^{4\varepsilon} \begin{cases}
    -20\pi^4 \left(\frac{1}{\varepsilon^2} + \frac{2}{\varepsilon} + 4 \right) + \mathcal{O}(\varepsilon^1), \ \ \ &d=2, \\
    2^{5}5\pi^6 \left(\frac{1}{3\varepsilon} + \frac{4}{9}(4- 3 \log 2) \right) + \mathcal{O}(\varepsilon^1), \ \ \  &d=3, \\
    \end{cases} \\
   H_{4}^{(2)} &= (2R)^{4\varepsilon} \begin{cases}
     4\pi^4, \ \ \ &d=2, \\
     \mathcal{O}(\varepsilon^1), \ \ \ &d=3, \\
    \end{cases} \\
     H_{4}^{(3)} &= (2R)^{4\varepsilon} \begin{cases}
       -2\pi^4\left(\frac{5}{\varepsilon^2} + \frac{2(5 + 3\log 2)}{\varepsilon} + 2 \log^2 2 + 12\log 2 + 20 \right) + \mathcal{O}(\varepsilon^1), \ \ \ &d=2, \\
    2^3\pi^6\left(\frac{10}{3\varepsilon} + \frac{1}{9}(178 - 9\pi - 84\log2)  \right) + \mathcal{O}(\varepsilon^1), \ \ \ &d=3. \\
     \end{cases} \\
  \end{aligned}
\end{equation}
%Substituting these values and the renormalized $\lambda$ using \eqref{eq:renormLambda} into \eqref{eq:FNLquartic}, we find \eqref{eq:tildeFNLquartic3D} and \eqref{eq:cNLquartic}.
%Substituting these values into \eqref{eq:FNLquartic}, expanding $\lambda_0$ to third order in the renormalized $\lambda$ using \eqref{eq:renormLambda}, and evaluating at the fixed point (to order $\varepsilon^2$) of $\lambda$, we find \eqref{eq:tildeFNLquartic3D} and \eqref{eq:cNLquartic}.
To find \eqref{eq:tildeFNLquartic3D} and \eqref{eq:cNLquartic}, which are fourth order in $\varepsilon$, one uses the above values of $H_2$, $H_3$, and $H_4$ to first expand \eqref{eq:FNLquartic} to fourth order in the renormalized $\lambda$, which is found by plugging \eqref{eq:renormLambda} into \eqref{eq:lambdaCT}. 
Then the $1/\varepsilon$ poles cancel and in $d=3$ we get
\begin{equation}
  \begin{aligned}
    \widetilde{F}^{\rm LR} &= N \widetilde{F}_{\rm free}^{\rm LR} - \frac{N(N+2)}{144(4\pi)^2}\left( 3 \varepsilon + \alpha \varepsilon^2\right)\lambda^2  + \frac{N(N+2)(N+8)}{9(4\pi)^4}\left(1 + \alpha \varepsilon\right)\lambda^3 \\
    &\quad - \frac{4N(N+2)}{9(4\pi)^6}\left( (6-3\pi+12\log(2))(5N+22) +(N+8)^2\alpha \right) \lambda^4 + \cdots,
  \end{aligned}
\end{equation}
where $\alpha = 8 - 3\pi + 6\log(4e^{\gamma_E}\mu R)$ and the dots denote terms that will be $\mathcal{O}(\varepsilon^5)$ when evaluated at the fixed point $\lambda_*$. After substituting $\lambda_*$ found by solving $\beta(\lambda_*) = 0$ to order $\varepsilon^2$, we find that the logarithmic terms cancel, leading to (\ref{eq:tildeFNLquartic3D}). The $d=2$ case works analogously.  
%Second, one then evaluates this fourth-order-in-$\lambda$ expression at the fixed point $\lambda_*$ found by solving $\beta(\lambda_*) = 0$ to order $\varepsilon^2$ using \eqref{eq:betaLambda}. 

\subsection{Cubic model} \label{app:NLcubic}

The diagrams correcting the three-point vertex were considered in momentum-space in \cite{Fei:2014yja}. Here, we evaluate them in the long-range model. The one-loop corrections of the cubic vertices, including the counterterms, take the following form: 
\begin{align}
    -(g_1+\delta_{g_1} )-(g_1+g_2)g_1^2 I, \quad -(g_2+\delta_{g_2})-(Ng_1^3+g_2^3)I,
\end{align}
where $I$ is the momentum space integral 
\begin{align}
    I = \int\frac{d^d k}{(2\pi)^d}\frac{1}{|k|^s |p-k|^s |k+q|^s}~.
\end{align}
Here we have used the momentum space representation $G(p) = \frac{1}{|p|^s}$ of the free propagator in the long-range model. We evaluate the integral $I$ by using Feynman parameterization: 
\begin{equation}
\begin{split}
    &I= \int_0^1\frac{dx dy dz}{ x y z}\frac{\delta(1-x-y-z) (x y z )^{\frac{s}{2}}}{(4\pi)^{\frac{d}{2}}\Gamma(\frac{s}{2})^3}\frac{\Gamma(\frac{3s-d}{2})}{\Delta^{\frac{3s-d}{2}}},\\
    &\Delta = x(1-x)p^2+y(1-y)q^2+2 xy p\cdot q~.
    \end{split}
\end{equation}
Because $3s-d=2\varepsilon$, the integral $I$ has a $1/\varepsilon$ pole:  
\begin{align}
     I  = \frac{1}{(4\pi)^{\frac{d}{2}}\Gamma(\frac{d}{2})\, \varepsilon} + \mathcal{O}(\varepsilon^0).
\end{align}
From this divergence, we can directly find the leading-order beta functions \eqref{eq:betagNL}. The free energy \eqref{eq:FreeEnergyNLcubic} involves the diagrams
\begin{align}
    & G_2=3!t_2C_{d,s}^3I_2\left(d-\varepsilon \right),\\
    & G_4=3(3!)^3 C_{d,s}^6(3t_{41}G_4^{(1)}+2t_{42} G_4^{(2)}),
\end{align}
where the $t$ coefficients are given by \eqref{eq:tcoefs}. $G_4^{(1)}$ (corresponding to the diagram $\begin{tikzpicture} [baseline={([yshift=-.5ex]current bounding box.center)}]
\draw (-0.2, -0.15)node[vertex]{} to (-0.2, 0.15)node[vertex]{} ;
\draw (0.2, -0.15)node[vertex]{} to (0.2, 0.15)node[vertex]{} ;
\draw (-0.2, 0.15) to [in=150, out=30](0.2, 0.15) to [in=-30, out=-150] (-0.2, 0.15);
\draw (-0.2, -0.15) to [in=150, out=30](0.2, -0.15) to [in=-30, out=-150] (-0.2, -0.15);
\end{tikzpicture} $) is of order $\varepsilon$ in odd dimensions and is $\mathcal{O}(1)$ in even dimensions,\footnote{The finiteness of this diagram is closely related to the non-renormalization property of $\phi$ and $\sigma$ in the long-range model.} e.g. 
\begin{equation}
  \begin{aligned}
    G_4^{(1)} &= (2R)^{4\varepsilon} \begin{cases}
      9 \pi^4 + \mathcal{O}(\varepsilon^1),&d=2, \\
       -\frac{27\pi^8}{80} + \mathcal{O}(\varepsilon^1),&d=4.
    \end{cases}
  \end{aligned}
\end{equation}
$G_4^{(2)}$  (corresponding to the diagram $\begin{tikzpicture} [baseline={([yshift=-.5ex]current bounding box.center)}]
\draw (-0.2, -0.15)node[vertex]{} to (-0.2, 0.15)node[vertex]{} ;
\draw (0.2, -0.15)node[vertex]{} to (0.2, 0.15)node[vertex]{} ;
\draw (-0.2, 0.15) to  (0.2, -0.15);
\draw (-0.2, -0.15) to (0.2, 0.15);
\draw (-0.2, 0.15) to  (0.2, 0.15);
\draw (-0.2, -0.15) to (0.2,- 0.15);
\end{tikzpicture} $)  is $\mathcal{O}(1)$ in odd $d$, e.g.
\begin{equation}
  \begin{aligned}
    G_4^{(2)} &= (2R)^{4\varepsilon}  \begin{cases}
      4 \pi^8 + \mathcal{O}(\varepsilon^1),&d=3, \\
       -\frac{9\pi^{12}}{10\Gamma(\frac{1}{3})^3} + \mathcal{O}(\varepsilon^1),&d=5. 
      \end{cases}
  \end{aligned}
\end{equation}
and contains a $1/\varepsilon$ pole in even $d$, e.g.
\begin{equation}
  \begin{aligned}
    G_4^{(2)} &= (2R)^{4\varepsilon}  \begin{cases}
      - \frac{3}{4} \pi^{5/2} \Gamma(\frac{1}{6})^3\left(\frac{1}{\varepsilon} + 2\right) + \mathcal{O}(\varepsilon^1),&d=2, \\
      \frac{27\pi^{19/2}}{\Gamma(\frac{1}{6})^3}\left(\frac{1}{\varepsilon} + 3 \right) + \mathcal{O}(\varepsilon^1) ,&d=4.
      \end{cases}
  \end{aligned}
\end{equation}
Then, including terms that will be of order $\varepsilon^2$ at the fixed point of \eqref{eq:betagNL}, and using the definition \eqref{eq:LRdef}, we find \eqref{eq:deltaFLRgen}.

\bibliographystyle{JHEP}
\bibliography{refs.bib}

\end{document}